\documentclass[11pt]{article}

\usepackage{latexsym}
\usepackage{amssymb}
\usepackage{eurosym}
\usepackage{graphicx}
\usepackage{float}
\usepackage[small]{caption}
\usepackage{subfig}
\usepackage{psfrag}
\usepackage[english]{babel}
\usepackage[utf8x]{inputenc}
\usepackage{ucs}
\usepackage{amsmath}
\usepackage{amsfonts}

\newtheorem{Theorem}{Theorem}[part]

\newtheorem{Proposition}{Proposition}[part]

\newtheorem{Remark}{Remark}[part]

\def \Frac{\displaystyle\frac}

\def \Sup{\displaystyle\sup}

\def \Liminf{\displaystyle\liminf}

\def \Limsup{\displaystyle\limsup}

\def \Min{\displaystyle\min}

\def \N{\mathbb{N}}
\def \R{\mathbb{R}}

\def \Z{\mathbb{Z}}
\def \E{\mathbb{E}}
\def \F{\mathbb{F}}

\def \P{\mathbb{P}}

\def \T{\mathbb{T}}
\def \X{\mathbb{X}}
\def \Y{\mathbb{Y}}
\def \Z{\mathbb{Z}}

\def \TT{{\bf T}}

\def \Ac{{\cal A}}

\def \Cc{{\cal C}}

\def \Ec{{\cal E}}
\def \Fc{{\cal F}}
\def \Gc{{\cal G}}
\def \Hc{{\cal H}}

\def \Lc{{\cal L}}

\def \Sc{{\cal S}}

\def \Zc{{\cal Z}}

\def \trib{\blacktriangleright}

\def \eps{\varepsilon}

\def \ep{\hbox{ }\hfill$\Box$}

\def\Dt#1{\Frac{\partial #1}{\partial t}}

\def\Dp#1{\Frac{\partial #1}{\partial p}}

\def\Dpp#1{\Frac{\partial^2 #1}{\partial p^2}}

\def\Dpp#1{\Frac{\partial^2 #1}{\partial p^2}}

\def \Frac{\displaystyle\frac}

\def \Sup{\displaystyle\sup}

\def \Liminf{\displaystyle\liminf}
\def \Limsup{\displaystyle\limsup}

\def \Min{\displaystyle\min}

\def\reff#1{{\rm(\ref{#1})}}

\def\beqs{\begin{eqnarray*}}
\def\enqs{\end{eqnarray*}}
\def\beq{\begin{eqnarray}}
\def\enq{\end{eqnarray}}

\def\reff#1{{\rm(\ref{#1})}}

\begin{document}

\title{Numerical methods for an optimal order execution problem \thanks{We would like to thank J.G. Grebet (EXQIM) for  discussions and  remarks  during the preparation of this work.}}

\author{Fabien Guilbaud\footnote{EXQIM, and LPMA, University Paris 7, fabien.guilbaud@exqim.com}~~~
Mohamed Mnif\footnote{ENIT, Tunis,  mohamed.mnif@enit.rnu.tn}~~~
Huy\^en Pham\footnote{LPMA, University Paris 7, CREST-ENSAE, and Institut Universitaire de France, pham@math.jussieu.fr}
}


\maketitle

\begin{abstract}
This paper  deals with numerical solutions to an impulse control problem arising from optimal 
portfolio liquidation with  bid-ask spread and  market price impact pena\-lizing speedy execution trades.  
The corresponding dynamic programming (DP) equation is a quasi-variational inequality (QVI)  with solvency constraint satisfied by the value function in the sense of constrained viscosity solutions.  By taking advantage of the lag variable tracking the time interval between trades, 
we can provide an explicit backward numerical scheme for the time discretization of the DPQVI. The convergence of this discrete-time scheme is shown by viscosity solutions arguments. An optimal quantization method  is used for computing  the (conditional) expectations arising in this scheme.  Numerical results are presented by examining the behaviour of optimal liquidation strategies, and comparative performance analysis with respect to some benchmark execution strategies.  We also illustrate our optimal liquidation algorithm on real data, and observe various interesting  patterns of order execution  strategies. Finally, we provide some numerical tests of sensitivity with respect to the bid/ask spread and market impact para\-meters.
\end{abstract}


\vspace{7mm}

\noindent {\bf JEL Classification~:} G11.

\vspace{3mm}

\noindent {\bf MSC Classification (2000)~:} 93E20, 65C05, 91B28, 60H30.

\newpage

\section{Introduction}

\setcounter{equation}{0} 
\setcounter{Assumption}{0} 
\setcounter{Example}{0} 
\setcounter{Theorem}{0} 
\setcounter{Proposition}{0}
\setcounter{Corollary}{0} 
\setcounter{Lemma}{0} 
\setcounter{Definition}{0} 
\setcounter{Remark}{0}

Portfolios managers define ``implementation shortfall" as the difference in performance bet\-ween a theoretical trading strategy and the implemented portfolio. In a theoretical strategy, the investor observes  price displayed by the market and assumes that trades will actually be executed at this price. Implementation shortfall measures the distance between the realized transaction price and the pre-trade decision price. Indeed, the investor has to face several adverse effects when executing a trading strategy, usually referred to as trading costs. Let us describe the three main components of these illiquidity effects: the bid/ask spread, the broker's fees and the market impact. The best bid  (resp. best ask) price is the best offer to buy (resp. to sell) the asset, and the bid/ask spread is the difference (always positive in the continuous trading session) between the best ask price and  best bid price. The broker's fees are the amount paid to the broker for executing the order. The market impact refers to the following phenomenon: any buy or sell market order passed by an investor induces an adverse market reaction that will penalize quoted price from the investor point of view. 

Market impact is a key factor when executing large orders. A famous  worst case example is  J\'erome Kerviel's liquidation portfolio, operated by Soci\'et\'e  G\'en\'erale in 2008. According to the report of \textit{Commission Bancaire}, the liquidative value of Kerviel's portfolio was -2,7G\geneuro{} when the Soci\'et\'e  G\'en\'erale decided to unwind it on January 20, 2008. The liquidation was operated during 3 days and led to a supplementary loss of 3,6G\geneuro{}. Even in regular operations, price impact may noticeably affect a trading strategy. On April 29, 2010, Reuters agency reports that Citadel Investment Group sold 170M shares of the E*Trade stock, and raised about 301M\$: this operation led to a price fall of 7,1\%. These examples explain why measurement and efficient management of market impact is a key issue for financial institutions, and the research of low-touch trading strategies has found a great interest among academics.

Most of market places and brokers offer  several common tools to reduce market impact. We can cite as an example the simple time slicing (we will refer to this example later as the \textit{uniform strategy}): a large order is split up in multiple \textit{children} orders of the same size, and these children orders are sent to the market at regular time intervals. Brokers also propose more sophisticated tools as \textit{smart order routing} (SOR) or volume weighted average price (VWAP) based algorithmic strategies. 
Indeed, one basic observation is that market impact can be reduced by splitting up a large order into several children orders. Then the investor has to face the following trade-off: if he chooses to trade immediately, he will penalize his performance due to market impact; if he trades gradually, he is exposed to price variation on the period of the operation. Our goal in this article is  to provide a numerical method to find optimal schedule and associated quantities for the children orders.

Recently, there has been considerable interest for this problem in the academic lite\-rature. The seminal papers  \cite{berlo98} and  \cite{almcri01} first provided a framework for managing market impact in a discrete-time model. The optimality is determined according to a mean-variance criterion, and this leads to a static strategy, in the sense that it is independent of the stock price. 
Models of market impact based on stylized order book dynamics were proposed in   \cite{obiwan05},   \cite{schsch07} and  \cite{gatsch10}. There also has been several optimal control approaches to the order execution problem, using a penalizing function to model price impact: 
the papers \cite{rogsin08} and  \cite{for09} assume  continuous-time trading, and 
use an Hamilton-Jacobi-Bellman approach for the mean-variance criterion, while \cite{hemam04},   \cite{lyvmnipha07}, and \cite{khapha09} 
consider real trading taking place in discrete-time by  using an impulse control  approach. This last approach combines the advantages of realistic modelling of portfolio liqui\-dation and the tractability of continuous-time stochastic calculus. In these papers, the optimal liquidation strategies are  price-dependent in contrast with static strategies. 

In this article, we adopt the model investigated in \cite{khapha09}. Let us describe the main features of this model. 
The stock price process is assumed to follow a geometrical Brownian motion. The price impact is modelled via a nonlinear transaction costs function, that depends both on the quantity traded, and on a lag variable $\theta$ tracking  the time spent since the investor's last trade. This lag variable will penalize rapid execution trades, and ensures 
in particular that trading times are strictly increasing, which is consistent with market practice  in limit order books. 
In this context, we consider the problem of an investor seeking to unwind an initial position in stock shares over a finite horizon.  
Risk aversion of the investor is modelled through a utility function, and we use an impulse control approach for 
the optimal order execution problem, which consists in maximizing the expected utility from terminal liquidation wealth, under a natural economic solvency constraint involving the liquidation value of portfolio.  The theoretical part of this impulse control problem is studied in \cite{khapha09}, and the solution is characterized through dynamic programming 
by means of  a  quasi-variational inequality (QVI) satisfied by the value function in the (constrained) viscosity sense. 
The aim of this paper is to solve numerically this optimal order execution problem. There are actually few papers dealing with a complete numerical treatment of impulse control problems, see  \cite{chaokssul02}, \cite{mar06}, or \cite{chefor08}. In these papers, the domain has a simple shape, typically  rectangular, and a finite-difference method is used.  In contrast, our domain is rather complex  due to the solvency constraint naturally imposed by the liquidation value under market impact, and we propose a suitable probabilistic numerical method for solving the associated impulse control problem. Our main contributions are the following: 
\begin{itemize}
\item We provide a numerical scheme for  the  QVI associated to the impulse control problem and prove that this method is monotone, consistent and stable, hence converges to the viscosity solution of the QVI. For this purpose, we adapt a proof from  \cite{barsou91}.
\item We take advantage of the lag variable $\theta$ to provide an explicit backward scheme and then simplify the computation of the solution. This contrasts with the classical approach by iterative sequence of optimal stopping problems, see e.g. \cite{chaokssul02}. 
\item We provide the detailed computational probabilistic algorithm with an optimal quantization method for the approximation of  conditional expectations arising in the backward scheme. 
\item We provide several numerical tests and statistics, both on simulated and real data, and compare the optimal strategy to a benchmark of two other strategies: the uniform strategy and the naive one consisting in the liquidation of all shares in one block  at the terminal date.  We also provide some sensitivity numerical analysis with respect to the bid/ask spread and market impact parameters. 
\end{itemize}
This paper is organized as follows: Section 2 recalls the problem formulation and main properties of the model, in particular the PDE characterization 
of the  impulse control problem by means of constrained viscosity solutions to the QVI,  as stated in \cite{khapha09}. Section 3 is devoted to the time discretization and the proof of convergence of the numerical scheme. Section 4 provides the numerical algorithm and numerical methods to solve the DPQVI. Section 5 presents the results obtained with our implementation, both on simulated and historical data.

\section{Problem formulation}

\setcounter{equation}{0} 
\setcounter{Assumption}{0} 
\setcounter{Example}{0} 
\setcounter{Theorem}{0} 
\setcounter{Proposition}{0}
\setcounter{Corollary}{0} 
\setcounter{Lemma}{0} 
\setcounter{Definition}{0} 
\setcounter{Remark}{0}

\subsection{The model of portfolio liquidation}

We consider a financial market  where an  investor has to  
liquidate an initial  position of $y$ $>$ $0$ shares of risky asset by time $T$.  He faces the following risk/cost tradeoff:  if  he trades rapidly, this results in higher costs due to market impact; if he   
splits the order into several smaller blocks, he is exposed to the risk of price depreciation during the trading horizon.  
 We adopt  the recent continuous-time framework of \cite{khapha09}, who proposed  a modeling where  trading takes place at discrete random times  through an impulse control formulation,  and with a temporary price  impact depending  on the time interval between trades, and including a bid-ask spread. 

Let us recall  the details of the model. We  set a probability space  $(\Omega,\Fc,\P)$  equipped with a filtration 
$\F$ $=$ $(\Fc_t)_{0\leq t\leq T}$ supporting a one-dimensional Brownian motion $W$ on a finite horizon $[0,T]$, $T$ $<$ $\infty$. 
We denote by $P_t$ the market price of the  risky asset, by $X_t$ the cash holdings, 
by $Y_t$ the number of stock shares held by the investor
at time $t$ and by $\Theta_t$ the time interval between $t$ and the last trade before $t$.

\vspace{1mm}

{\it Trading strategies.} We assume that the investor can only trade at discrete time on $[0,T]$.
This is modelled through an impulse control strategy $\alpha$  $=$
$(\tau_n,\zeta_n)_{n\geq 1}$ where  $\tau_1$ $\leq$ $\ldots$ $\tau_n$ $\leq$ $\ldots$ $\leq$ $T$ are stopping times
representing the trading times and $\zeta_n$, $n$ $\geq$ $1$, are
$\Fc_{\tau_n}$-measurable random variables valued in $\R$ and giving the quantity of stocks purchased if $\zeta_n$ $\geq$ $0$
or selled if  $\zeta_n$ $<$ $0$ at these times. A priori, the sequence $(\tau_n,\zeta_n)$ may be  finite or infinite. 
We introduce the lag variable tracking the time interval between trades, 
which evolves according to 
\beq\label{dynTheta}
\Theta_t = t-\tau_n ,\,\, \tau_n \leq t < \tau_{n+1},\; \;\,\,\,\Theta_{\tau_{n+1}}=0,\,\, n\geq 0.
\enq
The dynamics of the number of stock shares $Y$ is then given by~:
\beq
Y_s &=& Y_{\tau_n}, \;\;\;  \tau_n \leq s < \tau_{n+1} \label{dynY0}, \;\;\;\;\; Y_{\tau_{n+1}} \;=\; Y_{\tau_n} + \zeta_{n+1}, \;\;\; n \geq 0.  \label{dynY}
\enq

\vspace{1mm}

{\it Cost of illiquidity.} The market price of the risky asset  process  follows a geometric Brownian motion:
\beq \label{dynP0}
dP_t &=&  P_t (b dt + \sigma dW_t),
\enq
with constant $b$ and $\sigma$ $>$ $0$.  We do not consider a permanent price impact, i.e. the lasting effect of large trade, but focus here on the temporary price impact that penalize the price at which an investor will trade the asset.  Suppose now that the investor decides at time $t$ to trade the quantity $e$. If the current market price is $p$, and the time lag from the last order is $\theta$, then the price he actually get  for the order $e$ is: 
\beq \label{Qimpact}
Q(e,p,\theta) & = &  p f (e,\theta),
\enq
where  $f$ is a temporary price impact function from $\R\times [0,T]$ into $\R_{+}\cup\{\infty\}$.  
Actually, in the rest of the paper, we consider a  function $f$ in the form
\beq \label{Qexp}
f(e, \theta) &=&  \exp{\big(\lambda |\frac{e}{\theta}|^{\beta} {\rm sgn}(e)\big)}.
\big( \kappa_a{\bf 1}_{e>0}+{\bf 1}_{e=0} +\kappa_b{\bf 1}_{e<0} \big) ,
\enq
where $\beta>0$ is the price impact exponent, $\lambda$ $>$ $0$ is the temporary price impact factor, $\kappa_b$ $<$ $1$, and $\kappa_a$ $>$ $1$  are the bid and ask  spread parameters.  The impact of liquidity modelled in \reff{Qimpact} is like a transaction cost combining nonlinearity and proportionality effects. The  nonlinear costs come from  the dependence of the function $f$ on $e$, but also on  $\theta$. 
On the other hand, this transaction cost  function $f$  can be determined implicitly from the impact of a market order placed by a large trader in a limit order book, as explained in \cite{obiwan05}, \cite{schsch07} or \cite{rogsin08}.  Moreover, the  dependance of $f$ in $\theta$ in \reff{Qexp}  means that rapid trading has a larger temporary price impact than slower trading. Such kind of assumption is also made in the seminal paper \cite{almcri01}, and  reflects stylized facts on limit order books. 
The form \reff{Qexp} was suggested in several 
empirical studies, see \cite{lilfarman03}, \cite{potbou03}, \cite{almthuhau05}, and used also in \cite{for09}, \cite{khapha09}.

\vspace{1mm}

{\it Cash holdings.} 
We assume a zero risk-free return, so that the cash holdings are constant between two trading times: 
\beq \label{Xconst}
X_t &=& X_{\tau_n}, \;\;\;\; \tau_n \leq t  <  \tau_{n+1}, \;\; n \geq 0.
\enq
When a discrete trading $\Delta Y_{t}$ $=$ $\zeta_{n+1}$ occurs at time $t$ $=$ $\tau_{n+1}$,  
this results  in a variation of the cash amount given by $\Delta X_t$ $:=$ $X_t-X_{t^-}$ $=$ $-\Delta Y_t . Q(\Delta Y_t,P_t,\Theta_{t^-})$ 
due to the illiquidity effects.  In other words,  we have
\beq
X_{\tau_{n+1}}  
&=&  X_{\tau_{n+1}^-} - \zeta_{n+1}  P_{\tau_{n+1}} f( \zeta_{n+1}, \tau_{n+1}-\tau_{n} ), \;\;\; n \geq 0.  \label{X}  
\enq

\begin{Remark} \label{remtrade}
{\rm
Notice that since $f(e,0)=0$ if $e<0$ and $f(e,0)=\infty$ if $e>0$, an immediate sale does not increase the cash holdings, i.e. 
$X_{\tau_{n+1}}\,=\,X_{\tau_{n+1}^{-}}\,=\,X_{\tau_{n}}$, while an immediate purchase leads to a bankruptcy i.e. 
$X_{\tau_{n+1}}\,=\,-\infty$. 
}
\end{Remark}

\vspace{1mm}

{\it Liquidation value and solvency constraint.} 
The solvency constraint is a key issue in portfolio choice problem.
The point is to define in an economically meaningful  way what is the portfolio value of a position in cash and stocks.  
In our context, we first impose a no-short selling constraint on the trading strategies, i.e.
\beqs
Y_t &\geq& 0,\,\,\;\;\;  0\leq t\leq T.
\enqs
Next, we introduce the liquidation function $L(x,y,p,\theta)$ representing  the value that an investor would
obtain by liquidating immediately his stock position $y$  by a single block trade, when the pre-trade price is $p$ and the time lag from the last order is $\theta$. It is defined on $\R\times \R_+\times (0,\infty)\times [0,T]$ by
\beqs
L(x,y,p,\theta)=x+ypf(-y,\theta),
\enqs
and we constrain the portfolio's liquidative value to satisfy the solvency criterion:
\beqs
L(X_t,Y_t,P_t,\Theta_t) &\geq& 0,\,\,\;\;\; 0\leq t\leq T.
\enqs
We then naturally introduce the solvency  region:
\beqs
\Sc & = & \left\{ (z, \theta)=(x,y,p,\theta) 
\in \R\times\R_+\times (0,\infty)\times [0,T]~: L(z,\theta) > 0 \right\}.
\enqs
and we denote its boundary and its closure by
\beqs
\partial \Sc \; = \partial_y \Sc \cup  \partial_L \Sc \;  & \mbox{ and } &
\bar\Sc  \; = \;   \Sc \cup \partial \Sc.
\enqs
where
\beqs
\partial_y \Sc &\; = \;&  \left\{ (z, \theta)=(x,y,p,\theta) 
\in \R\times\R_+\times (0,\infty)\times [0,T]: \; y=0 \mbox{ and }x=L(z,\theta) \geq  0 \right\},\\
\partial_L \Sc &\; = \;&  \left\{ (z, \theta)=(x,y,p,\theta) 
\in \R\times\R_+\times (0,\infty)\times [0,T]: \; L(z,\theta) =  0 \right\}.
\enqs
In the sequel, we also introduce the corner lines in $\partial\Sc$~:
\beqs
D_0 \; = \;  \{(0,0)\}\times (0,\infty)\times [0,T]\;=\; \partial_y\Sc \cap \partial_L\Sc.
\enqs

\vspace{1mm}

{\it Admissible trading strategies.}  Given  $(t,z,\theta)$ $\in$ $[0,T]\times\bar\Sc$,  we say that the impulse control strategy 
$\alpha$ $=$ $(\tau_{n},\zeta_{n})_{n\geq 0}$   is admissible, denoted by $\alpha$ $\in$ $\Ac(t,z,\theta)$,  
if $\tau_0$ $=$ $t-\theta$, $\tau_n$ $\geq$ $t$,  $n$ $\geq$ $1$,  and  the process 
$\{(Z_{s},\Theta_s) = (X_{s,}Y_{s},P_{s},\Theta_{s}), t\leq s\leq T\}$  solution to  \reff{dynTheta}-\reff{dynY}-\reff{dynP0}-\reff{Xconst}-\reff{X}, with  an initial state $(Z_{t^-},\Theta_{t^-})$ $=$ $(z,\theta)$ (and the convention that  $(Z_{t},\Theta_{t})$ $=$ $(z,\theta)$ if $\tau_1$ $>$ $t$),      
satisfies  $(Z_{s},\Theta_s)$ $\in$ $[0,T]\times\bar\Sc$ for all $s$ $\in$ $[t,T]$.  As usual, to  alleviate notations, we omit 
 the dependence of $(Z,\Theta)$ in  $(t,z,\theta,\alpha)$, when there is no ambiguity.

\vspace{1mm}

{\it Portfolio liquidation problem.} We consider a utility function $U$ from $\R_+$ into $\R$, strictly increasing, concave and w.l.o.g. $U(0)$ $=$ $0$,
and s.t. there exists  $K$ $\geq$ $0$, $\gamma$ $\in$ $[0,1)$~:
\beqs \label{growthU}
U(w) & \leq &  K w^\gamma, \;\;\; \forall w \geq 0.
\enqs
The problem of optimal portfolio liquidation  is formulated as
\beq \label{defvliquid}
v(t,z,\theta) &=&   \sup_{\alpha\in\Ac_\ell(t,z,\theta)} \E\big[ U(X_T) \big], \;\;\; (t,z,\theta) \in [0,T]\times\bar\Sc,
\enq
where   $\Ac_\ell(t,z,\theta)$ $=$ $\big\{\alpha \in \Ac(t,z,\theta):~ Y_{T}~=~0 \big\}$. As observed in \cite{khapha09}, one  can shift  the terminal liquidation constraint in $\Ac_\ell(t,z,\theta)$ to a terminal liquidation utility  by considering the function $U_L$  defined on $\bar \Sc$ by: 
\beqs
U_{L}(z,\theta) & = & U(L(z,\theta)), \;\;\; (z,\theta) \in \bar\Sc. 
\enqs
Then, problem \reff{defvliquid} is written equivalently in 
\beq \label{defbarv}
v(t,z,\theta) &=& \sup_{\alpha\in\Ac(t,z,\theta)} \E\Big[ U_L(Z_T,\Theta_T) \Big], \;\;\; (t,z,\theta) \in [0,T]\times\bar\Sc. 
\enq

\subsection{PDE characterization}

 The dynamic programming Hamilton-Jacobi-Bellman (HJB) equation corresponding to the stochastic control problem \reff{defvliquid} is a 
 quasi-variational inequality written as
\beq \label{HJB} 
\min\big[-\Dt{v} - \Lc v  \; , \; v - \Hc v \big] &=& 0, \;\;\;  \mbox{ on } \;\; [0,T)\times\bar \Sc, 
\enq 
together with the relaxed terminal condition 
\beq \label{HJBt} 
\min\left[ v- U_L, v - \Hc v \right] &=&0, \;\;\;  \mbox{ on } \;\;\{T\}\times \bar \Sc.
\enq 
Here, $\Lc$ is  the infinitesimal generator associated to the process $(Z=(X,Y,P),\Theta)$ in  a no-trading period: 
\beqs 
\Lc \varphi &=& \frac{\partial \varphi}{\partial \theta} + bp \Dp{\varphi} + \frac{1}{2}\sigma^2 p^2 \Dpp{\varphi}, 
\enqs 
$\Hc$ is the impulse operator defined by 
\beqs \Hc \varphi(t,z,\theta) &=&
\sup_{e\in\Cc(z,\theta)} \varphi(t,\Gamma(z,\theta,e),0), \;\;\; (t,z,\theta) \in [0,T]\times\bar\Sc, 
\enqs 
$\Gamma$ is the impulse transaction function defined from $\bar\Sc\times\R$ into  $\R\times\R\times(0,\infty)$: 
\beqs 
\Gamma(z,\theta, e) &=& (x-e p f(e,\theta),y+e,p), \;\; z =(x,y,p) \in \bar\Sc, \; e \in \R, 
\enqs 
and $\Cc(z,\theta)$ the set of admissible transactions~: 
\beqs 
\Cc(z,\theta) &=& \left\{ e \in \R~:
\Big(\Gamma(z,\theta,e),0\Big) \; \in \; \bar\Sc \right\}.  
\enqs 

By standard arguments,  we derive  the constrained viscosity solution property of the value function $v$ to  \reff{HJB}-\reff{HJBt}. 
However, in order to have a complete characterization of the value function via its HJB equation, we need a uniqueness result. 
Unfortunately, in our model, it seems not possible to get such result, at least by  classical arguments since there is no strict supersolution 
to \reff{HJB}. In \cite{khapha09}, the authors prove a weaker characterization of the value function in terms of minimal  solution to its HJB equation. 
They also  consider a small variation of the original model by  adding a fixed transaction fee $\eps$ $>$ $0$ 
at each trading. This means that given 
a trading strategy $\alpha$ $=$ $(\tau_n,\zeta_n)_{n\geq 0}$,  the controlled state process  $(Z=(X,Y,P),\Theta)$ jumps now at time 
$\tau_{n+1}$,  by: 
\beq \label{jumpeps}
(Z_{\tau_{n+1}},\Theta_{\tau_{n+1}}) &=& \Big( \Gamma_\eps(Z_{\tau_{n+1}^-},\Theta_{\tau_{n+1}^-},\zeta_{n+1}), 0 \Big), 
\enq
where $\Gamma_\eps$ is the function defined on $\R\times\R_+\times (0,\infty) \times [0,T]\times\R$ into 
$\R\cup\{-\infty\}\times\R\times (0,\infty)$ by:
\beqs
\Gamma_\eps(z,\theta,e) &=& \Gamma(z,\theta,e) - (\eps,0,0) \; = \; 
\Big( x - e p f(e,\theta) - \eps, y+ e, p\Big),
\enqs
for $z$ $=$ $(x,y,p)$. The dynamics of $(Z,\Theta)$ between trading dates is given as before. 
We introduce a modified liquidation function $L_\eps$ defined  by: 
\beqs
L_{\eps}(z,\theta) & = & \max[ x ,L(z,\theta) - \eps ], \;\;\; (z,\theta)  = (x,y,p,\theta) \in \R\times\R_+\times(0,\infty)\times [0,T]. 
\enqs 
The interpretation of this modified liquidation function is the following. Due to the presence of the transaction fee at each trading,  it may be advantageous for the investor not to liquidate his position in stock shares (which would give him $L(z,\theta)-\eps$), and rather bin his stock shares, 
by keeping only his cash amount (which would give him $x$).  Hence, the investor chooses 
 the best of these two possibilities, which induces a liquidation  value $L_\eps(z,\theta)$. 

The corresponding   solvency region $\Sc_\eps$ $\subset$ $\Sc$ 
with its closure $\bar\Sc_\eps$ $=$ $\Sc_\eps$ $\cup$ $\partial\Sc_\eps$, and boundary 
$\partial\Sc_\eps$ $=$ $\partial_y\Sc_\eps$ $\cup$ $\partial_L\Sc_\eps$ are given by:
\beqs
\Sc_{\eps} & = & \Big\{  (z,\theta) =(x,y,p,\theta)\in\R\times\R_{+}\times (0,\infty)\times [0,T]:~ y  > 0 \; \mbox{ and } L_{\eps}(z,\theta)>0  \Big\}, \\
\partial_y\Sc_\eps &=& \Big\{  (z,\theta) =(x,y,p,\theta)\in\R\times\R_{+}\times (0,\infty)\times [0,T]:~ y  = 0 \; \mbox{ and } 
\; L_{\eps}(z,\theta) \geq 0  \Big\}, \\
\partial_L\Sc_\eps &=& \Big\{  (z,\theta) =(x,y,p,\theta)\in\R\times\R_{+}\times (0,\infty)\times\R_{+}:~ L_{\eps}(z,\theta)=0  \Big\}. 
\enqs

The set of admissible trading strategies is defined as follows: 
given $(t,z,\theta)$ $\in$ $[0,T]\times\bar\Sc_\eps$, we say that the impulse control  $\alpha$ is admissible, denoted by $\alpha$ $\in$ 
$\Ac_\eps(t,z,\theta)$, if $\tau_0$ $=$ $t-\theta$, $\tau_n$ $\geq$ $t$, $n$ $\geq$ $1$,  and the controlled state process $(Z^\eps,\Theta)$  
solution to  \reff{dynTheta}-\reff{dynY}-\reff{dynP0}-\reff{Xconst}-\reff{jumpeps}, with  an initial state 
$(Z_{t^-}^\eps,\Theta_{t^-})$ $=$ $(z,\theta)$ (and the convention that  $(Z_{t}^\eps,\Theta_{t})$ $=$ $(z,\theta)$ if $\tau_1$ $>$ $t$),      
satisfies  $(Z_{s}^\eps,\Theta_s)$ $\in$ $[0,T]\times\bar\Sc_\eps$ for all $s$ $\in$ $[t,T]$.  Here, we stress the dependence of 
$Z^\eps$ $=$ $(X^\eps,Y,P)$ in  $\eps$ appearing in the transaction function  $\Gamma_\eps$, and we notice that it affects only the cash component.

The liquidation utility function in this model with  fixed transaction fee $\eps$  
is defined on $\bar\Sc_\eps$ by $U_{L_\eps}(z,\theta)$ $=$ $U(L_\eps(z,\theta))$, 
and the associated  optimal portfolio liquidation problem is defined via its value function by:
 \beq \label{defveps}
 v_{\varepsilon}(t,z,\theta) & = & \sup_{\alpha\in\Ac_\varepsilon(t,z,\theta)}\E\big[ U_{L_{\eps}}(Z_{T}^\eps,\Theta_T) \big], \;\;\; 
 (t,z,\theta) \in [0,T]\times \bar\Sc_\eps.
 \enq
 The dynamic programming equation associated to the control problem \reff{defveps} is
 \beq
 \min\Big[ -\Dt{v_{\varepsilon}} - \Lc v_{\varepsilon}~,~v_{\varepsilon} -\Hc_\eps v_{\varepsilon} \Big] & = & 0 
\;\;\;\; \mbox{ on } [0,T)\times\bar\Sc_\eps, \label{QVIvareps}  \\
\min\big[v_{\varepsilon}-U_{L_\epsilon},v_{\varepsilon} -\Hc_\eps v_\eps \big] & = & 0 \;\;\;\; \mbox{ on }  
\{T\}\times \bar\Sc_\eps, \label{termeps}
\enq
where   $\Hc_\eps$ is the impulse operator defined by
\beqs
\Hc_\eps w(t,z,\theta) &=& \sup_{e \in \Cc_\eps(z,\theta)} w(t,\Gamma_\eps(z,\theta,e),0), \;\;\; (t,z,\theta) \in [0,T]\times\bar\Sc_\eps,
\enqs
for any locally bounded function $w$ on $[0,T]\times\bar\Sc_\eps$, with the convention that $\Hc_\eps w(t,z,\theta)$ $=$ $-\infty$ when 
$\Cc_\eps(z,\theta)$ $=$ $\emptyset$, and  the set of admissible transactions in the model with fixed transaction fee is:
 \beqs
 \Cc_{\eps}(z,\theta) & = & \Big\{e\in\R: \Big(\Gamma_\eps(z,\theta,e),0\Big) \in \bar\Sc_\eps  \Big\}, \;\;\; (z,\theta) \in \bar\Sc_\eps.
 \enqs

 We recall from \cite{khapha09} that $v_\eps$ is in the set $\Gc( [0,T]\times \bar\Sc_{\epsilon})$ of  functions satisfying the growth condition: 
\beqs \label{growthveps}
 \Gc( [0,T]\times \bar\Sc_{\epsilon}) & = & \Big\{ \varphi:[0,T]\times \bar\Sc_{\epsilon}\longrightarrow \R \mbox{ s.t. } 
\Sup_{[0,T]\times \bar\Sc_{\epsilon}}\frac{|\varphi(t,z,\theta)|}{\big(1+ (x+yp)^{\gamma}\big)} < \infty  \Big\}. 
\enqs

 \begin{Remark} \label{vdec}
{\rm  The function $z$  $\rightarrow$ $v_\eps(t,z,0)$ is strictly increasing in the argument of cash holdings $x$, for 
$(z=(x,y,p),0)$ $\in$ $\bar\Sc_\eps$, and fixed $t$ $\in$ $[0,T]$. 
Indeed, for  $x$ $<$ $x'$, and $z$ $=$ $(x,y,p)$, $z'$ $=$ $(x',y,p)$, any strategy 
$\alpha$ $ \in$ $\Ac_\eps(t,z,\theta)$ with corresponding state process $(Z_s=(X_s,Y_s,P_s),\Theta_s)_{s\geq t}$, is also in $\Ac_\eps(t,z',\theta)$, and leads to an associated state process $(Z_s'=(X_s+x'-x,Y_s,P_s),\Theta_s)_{s\geq t}$.  Using the fact that the utility function is strictly increasing, we deduce that $v_\eps(t,x,y,p,0)$ $<$ $v_\eps(t,x',y,p,0)$.  Moreover, the function $z$  $\rightarrow$ $v_\eps(t,z,0)$ is nondecreasing  in the argument of number of shares $y$.  Indeed,  fix  $z$ $=$ $(x,y,p)$, and $z'$ $=$ $(x,y',p)$ with $y$ $\leq$ $y'$.  Given any arbitrary $\alpha$ 
$=$ $(\tau_n,\zeta_n)_n$ $\in$ $\Ac_\eps(t,z,0)$, consider the strategy $\alpha'$ $=$ $(\tau_n',\zeta_n')$, starting from $(x,y',p)$ at time $t$, 
which consists in trading again immediately at time $t$ by selling $y'-y$ shares (which does not change the cash holdings, see Remark 
\ref{remtrade}), and then follow the same strategy than $\alpha$. The corresponding state  process satisfies 
$(Z'_s,\Theta_s')$ $=$ $(Z_s,\Theta_s)$ a.s. for $s$ $\geq$ $t$, and in particular $\alpha'$ $\in$ $\Ac_\eps(t,z',0)$,  together with 
$\E[U_{L_\eps}(Z_T',\Theta_T')]$ $=$ $\E[U_{L_\eps}(Z_T,\Theta_T)]$ $\leq$ $v(t,z',\theta)$.  
Since $\alpha$ is arbitrary in $\Ac_\eps(t,z,0)$, this shows that $v(t,x,y,p,0)$ $\leq$ $v(t,x,y',p,0)$.  
}
\end{Remark}

In the sequel, we shall denote by  $\Gc_+([0,T]\times\bar\Sc_\eps)$ the set of functions  $\varphi$ in $\Gc([0,T]\times\bar\Sc_\eps)$ such that 
$\varphi(t,x,y,p,0)$  is strictly increasing in $x$ and nondecreasing in $y$.

 \begin{Remark} \label{remnotrade}
 {\rm   Fix $t$ $\in$ $[0,T]$. For $\theta$ $=$ $0$, and $z$ $=$ $(x,y,p)$ s.t.  $(z,0)$ $\in$ $\bar\Sc_\eps$,  the set of admissible transactions  
 $\Cc_\eps(z,0)$ $=$ $[-y,0]$ (and  $\Gamma_\eps(z,0,e)$ $=$ $(x-\eps,y+e,p)$ for $e$ $\in$ $\Cc_{\eps}(z,0)$)  if $x$ $\geq$ $\eps$, and is empty otherwise.  Thus, $\Hc_\eps w(t,z,0)$ $=$ $\sup_{e\in [-y,0]} w(t,x-\eps,y+e,p,0)$ if $x$ $\geq$ $\eps$, and is equal to $-\infty$ otherwise. 
  This implies in particular that 
 \beq \label{infH}
 \Hc_\eps w(t,z,0) & <  & w(t,z,0),
 \enq
 for any  $w$ $\in$  $\Gc_+([0,T]\times\bar\Sc_\eps)$,  which is the case of $v_\eps$ (see Remark \ref{vdec}). 
 Therefore,  due to the market impact function $f$ in \reff{Qexp} penalizing rapid trades,   
 it is not optimal to trade again immediately right after some trade, i.e.  the optimal trading times are strictly increasing. 
 }
 \end{Remark}

A main result in  \cite{khapha09}  is to provide  a unique PDE characterization of the value functions $v_\eps$, $\eps$ $>$ $0$, and 
to prove that  the sequence $(v_\eps)_\eps$ converges to the original value function $v$ as $\eps$ goes to zero.

\begin{Theorem} \label{theomainapprox}
(1) The sequence $(v_\eps)_\eps$ is nonincreasing, and converges  pointwise on 
$[0,T]\times(\bar\Sc\setminus\partial_L\Sc)$ towards  $v$  as $\eps$ goes to zero, with $v_\eps$ $\leq$ $v$. 

\noindent (2)  For any  $\eps$ $>$ $0$, the value function $v_\eps$ is continuous on $[0,T)\times\Sc_\eps$, and is  the unique 
(in $[0,T)\times\Sc_\eps)$ constrained viscosity solution to \reff{QVIvareps}-\reff{termeps}, satisfying the growth condition in 
$\Gc([0,T]\times\bar\Sc_\eps)$, and the boundary condition: 
\beq
\lim_{(t',z',\theta')\rightarrow (t,z,\theta)} v_\eps(t',z',\theta') &=& v_\epsilon(t,z,\theta) \nonumber \\
& = & U(0), \;\;\;  \forall (t,z=(0,0,p),\theta) \in [0,T]\times D_0.  \label{bounveps}
\enq
\end{Theorem}

The rest of this paper is devoted to the numerical analysis and resolution of the QVI \reff{QVIvareps}-\reff{termeps}, which then provides an $\eps$-approximation of the original optimal portfolio liquidation problem \reff{defvliquid}.

\section{Time discretization and convergence analysis}

\setcounter{equation}{0} 
\setcounter{Assumption}{0} 
\setcounter{Example}{0} 
\setcounter{Theorem}{0} 
\setcounter{Proposition}{0}
\setcounter{Corollary}{0} 
\setcounter{Lemma}{0} 
\setcounter{Definition}{0} 
\setcounter{Remark}{0}

In this section, we fix $\eps$ $>$ $0$, and we study  time discretization of the QVI \reff{QVIvareps}-\reff{termeps} characterizing the value function 
$v_\eps$.   For a  time discretization step $h$ $>$ $0$ on  the interval $[0,T]$, let us  consider the following approximation scheme: 
\beq \label{scheme1}
S^h(t,z,\theta,v^h(t,z,\theta),v^h) &=& 0, \;\;\; (t,z,\theta) \in [0,T]\times\bar\Sc_\eps,
\enq
where $S^h$ $:$ $[0,T]\times\bar\Sc_\eps\times\R\times  \Gc_+([0,T]\times\bar\Sc_\eps)$ $\rightarrow$ $\R$ is defined by
\begin{eqnarray}
& & S^{h}(t,z,\theta, r, \varphi) \label{defSh} \\
&:=&
\left\{
\begin{array}{ll}
 \Min\Big[ r - \E\big[\varphi(t+h , Z^{0,t,z}_{t+h},\Theta^{0,t,\theta}_{t+h} ) \big]\; , \; r - \Hc_{\epsilon} \varphi (t,z,\theta)   \Big]
&\mbox{ if } t \in [0,T-h]  \nonumber \\
\Min\Big[ r - \E\big[\varphi(T , Z^{0,t,z}_{T},\Theta^{0,t,\theta}_{T} ) \big]\; , \; r - \Hc_{\epsilon} \varphi (t,z,\theta)   \Big]
&\mbox{ if } t \in (T-h,T)  \nonumber \\
\Min\Big[r -U_{L_\epsilon}(z,\theta) \;, \; r - \Hc_{\epsilon}  \varphi (t,z,\theta)  \Big] &\mbox{ if } t =T.  \nonumber
\end{array}
\right.
\end{eqnarray}
 Here, $(Z^{0,t,z},\Theta^{0,t,\theta})$ denotes the state process starting from $(z,\theta)$ at time $t$, and without any  
impulse control strategy: it is given by  
\beqs \label{defZ0}
\Big(Z_s^{0,t,z},\Theta_s^{0,t,\theta}\Big) &=& (x,y,P_s^{t,p},\theta+ s-t), \;\;\; s \geq t, 
\enqs
with $P^{t,p}$  the solution to \reff{dynP0} starting from $p$ at time $t$.  Notice that \reff{scheme1} is formulated as  a backward scheme for the solution $v^h$  through:
\beq
v^h(T,z,\theta) &=& \max\big[ U_{L_\epsilon}(z,\theta) \; , \;   \Hc_{\epsilon}  v^h (T,z,\theta) \big],  \label{vhterm} \\
v^h(t,z,\theta) &=&  \max\Big[  \E\big[v^h(t+h , Z^{0,t,z}_{t+h},\theta + h) \big]\; , \Hc_{\epsilon} v^h (t,z,\theta) \Big], \;\;\;  0 \leq t  \leq T-h,
\label{vht}
\enq
and $v^h(t,z,\theta)$ $=$ $v^h(T-h,z,\theta)$ for $T-h<t<T$.  This approximation scheme seems a priori implicit due to the nonlocal obstacle 
term $\Hc_\eps$. This is typically the case in impulse control problems, and the usual  way (see e.g. \cite{chaokssul02}, \cite{mar06})
 to circumvent this problem is to iterate the scheme by considering a sequence of optimal stopping problems: 
\beqs
v^{h,n+1}(T,z,\theta) &=& \max\big[ U_{L_\epsilon}(z,\theta) \; , \;   \Hc_{\epsilon}  v^{h,n} (T,z,\theta) \big],   \\
v^{h,n+1} (t,z,\theta) &=&  \max\Big[  \E\big[v^{h,n+1}(t+h , Z^{0,t,z}_{t+h},\theta t+h ) \big]\; , \Hc_{\epsilon} v^{h,n} (t,z,\theta) \Big],
\enqs
starting from $v^{h,0}$ $=$ $\E[U_{L_\eps}(Z_T^{0,t,z},\Theta_T^{0,t,\theta})]$.   Here, we shall  
make the numerical scheme \reff{scheme1}  explicit, i.e. without iteration, by taking effect of the state variable $\theta$ in our model.  
Recall indeed  from Remark \ref{remnotrade} that it is not optimal to trade again immediately right after some trade. Thus, for $v^h$ $\in$ $\Gc_+([0,T]\times\bar\Sc_\eps)$, and any $(z',0)$ $\in$ $\bar\Sc_\eps$, 
we have from \reff{infH} and \reff{vhterm}-\reff{vht}: 
\beqs
v^h(T,z',0) &=&  U_{L_\epsilon}(z',0) \\
v^h(t,z',0) &=&  \E\big[v^h(t+h , Z^{0,t,z'}_{t+h}, h ) \big]. 
\enqs
Therefore, by using again the definition of $\Hc_\eps$ in the relations \reff{vhterm}-\reff{vht}, we see that the scheme \reff{scheme1} is written equivalently as 
an explicit backward scheme:
\beq
v^h(T,z,\theta) &=& \max\big[ U_{L_\epsilon}(z,\theta) \; , \;   \Hc_{\epsilon} U_{L_\epsilon}(z,\theta)\big],  \label{vhterm2} \\
v^h(t,z,\theta) &=&  \max\Big[  \E\big[v^h(t+h , Z^{0,t,z}_{t+h},\theta + h) \big]\; , \; 
\sup_{e\in\Cc_\eps(z,\theta)} \E\big[v^h(t+h , Z^{0,t,z_\theta^e}_{t+h}, h ) \big]   \Big], 
\label{vht2}
\enq
for $0 \leq t  \leq T-h$, and $v^h(t,z,\theta)$ $=$ $v^h(T-h,z,\theta)$ for $T-h<t<T$, 
where we denote $z_\theta^e$ $=$ $\Gamma_\eps(z,\theta,e)$ in \reff{vht2}  to alleviate notations.  Notice that at this stage, this approximation scheme is not yet fully implementable since it requires   an approximation method for  the  expectations arising in \reff{vht2}.  
This is the concern of the next section.

We focus now on the convergence (when $h$ goes to zero)  of the solution $v^h$ to  \reff{scheme1} 
towards the value function $v_\eps$ solution to \reff{QVIvareps}-\reff{termeps}. Following \cite{barsou91},  we have to show that the scheme 
$S^h$ in \reff{defSh}  satisfies monotonicity, stability and consistency properties.  As usual, the monotonicity property  follows directly from  the definition \reff{defSh} of the scheme.

 \begin{Proposition} (Monotonicity) 

\noindent For all $h$ $>$ $0$, $(t,z,\theta)$  $\in$ $[0,T]\times \bar\Sc_{\epsilon}$, $r$ $\in$ $\R$, and $\varphi$, $\psi$ $\in$ 
$\Gc_+( [0,T]\times {\bar\Sc}_{\epsilon})$ s.t. $\varphi$ $\leq$ $\psi$, we have
\begin{eqnarray*}
S^{h}(t,z,\theta,r,\varphi) & \geq &  S^{h}(t,z,\theta,r,\psi).
\enqs
\end{Proposition} 

\vspace{1mm}

We next prove the stability property.    

\begin{Proposition} (Stability) 

\noindent For all $h$ $>$ $0$, there exists a  unique solution $v^h$ $\in$ 
$\Gc_+([0,T]\times\bar\Sc_\eps)$ to \reff{scheme1}, and the sequence $(v^h)_h$ is uniformly bounded in $\Gc([0,T]\times\bar\Sc_\eps)$: there exists 
$w$ $\in$  $\Gc([0,T]\times\bar\Sc_\eps)$ s.t. $|v^h|$ $\leq$ $|w|$ for all $h$ $>$ $0$. 
\end{Proposition}
{\bf Proof.} The uniqueness of a solution $\in$ $\Gc_+([0,T]\times\bar\Sc_\eps)$  to \reff{scheme1} follows from the explicit backward scheme 
\reff{vhterm2}-\reff{vht2}. For $t$ $\in$ $[0,T]$, denote by  $N_{t,h}$ the integer part of $(T-t)/h$, and $\TT_{t,h}$ $=$ 
$\{t_k=t+ kh, k=0,\ldots,N_{t,h}\}$ the partition of the interval $[t,T]$ with time step $h$. 
For $(t,z,\theta)$ $\in$ $[0,T]\times\bar\Sc_\eps$,  we denote by $\Ac_\eps^h(t,z,\theta)$ the subset of elements $\alpha$ $=$ 
$(\tau_n,\zeta_n)_n$ in $\Ac_\eps(t,z,\theta)$ such that the trading times $\tau_n$ are valued in $\TT_{t,h}$. 
Let us then consider the impulse control problem
\beq \label{defvh}
v^h(t,z,\theta) & = & \sup_{\alpha\in\Ac_{\varepsilon}^h(t,z,\theta)}\E\big[ U_{L_{\eps}}(Z_{T}^\eps,\Theta_T) \big], \;\;\; 
(t,z,\theta) \in [0,T]\times \bar\Sc_\eps. 
 \enq
It is clear from the representation \reff{defvh} that for all  $h$ $>$ $0$,  $0$ $\leq$ $v^h$ $\leq$ $v_\eps$, which shows that 
the sequence $(v^h)_h$ is uniformly bounded in $\Gc([0,T]\times\bar\Sc_\eps)$. 
Moreover, similarly as for $v_\eps$, and by the same arguments as in 
Remark \ref{vdec}, we see that $v^h(t,z,0)$ is strictly increasing in $x$ and nondecreasing in $y$ for $(z,0)$ $=$ $(x,y,p,0)$ $\in$ 
$\bar\Sc_\eps$.  Finally, we  observe that the numerical scheme \reff{scheme1} is  
the dynamic progra\-mming equation satisfied by the value function $v^h$. This  proves the required stability result.  
\ep

\vspace{2mm}

We now  move on the consistency property.

\begin{Proposition} (Consistency) 
 
\noindent   (i) For all $(t,z,\theta)\in [0,T)\times \bar\Sc_{\epsilon}$ and $\phi\in C^{1,2}([0,T)\times\bar\Sc_{\epsilon})$, we have
\begin{eqnarray}\label{defconsistent1}
& \Limsup_{\tiny{\stackrel{(h,t^{'},z^{'},\theta^{'})\rightarrow(0,t,z,\theta)}{ (t^{'},z^{'},\theta^{'})\in [0,T)\times \Sc_{\epsilon}}}}&
\Min\left\{\frac {\phi(t^{'},z^{'},\theta^{'})
- \E\Big[\phi(t^{'}+h , Z^{0,t^{'},z^{'}}_{t^{'}+h},\Theta^{0,t^{'},\theta^{'}}_{t^{'}+h} )\Big] }{h}, \Big(\phi-\Hc_{\epsilon}\phi\Big)(t^{'},z^{'},\theta^{'})  \right\}\nonumber\\
&\leq &\Min\Big\{ \Big(-\Dt{\phi} - \Lc \phi\Big)(t,z,\theta),\Big(\phi-\Hc_{\epsilon}\phi\Big)(t,z,\theta)
 \Big\}
\end{eqnarray}
and 
\begin{eqnarray}\label{defconsistent2}
& \Liminf_{\small{\stackrel{(h,t^{'},z^{'},\theta^{'})\rightarrow(0,t,z,\theta)}{ (t^{'},z^{'},\theta^{'})\in [0,T)\times \Sc_{\epsilon}}}}&
\Min\left\{\frac {\phi(t^{'},z^{'},\theta^{'})
- \E\Big[\phi(t^{'}+h , Z^{0,t^{'},z^{'}}_{t^{'}+h},\Theta^{0,t^{'},z^{'}}_{t^{'}+h} )\Big] }{h}, \Big(\phi-\Hc_{\epsilon}\phi\Big)(t^{'},z^{'},\theta^{'})  \right\}\nonumber\\
&\geq &\Min\Big\{\Big(-\Dt{\phi} - \Lc \phi\Big)(t,z,\theta), \Big(\phi-\Hc_{\epsilon}\phi\Big)(t,z,\theta)
 \Big\}
\end{eqnarray}

\noindent (ii)  For all  $(z,\theta)\in \bar\Sc_{\epsilon}$ and $\phi\in C^{1,2}([0,T]\times\bar\Sc_{\epsilon})$, we have
\begin{eqnarray}\label{defconsistent3}
& \Limsup_{\stackrel{(t^{'},z^{'},\theta^{'})\rightarrow(T,z,\theta)}{ (t^{'},z^{'},\theta^{'})\in [0,T)\times \Sc_{\epsilon}}}&
\Min\Big\{\phi(t^{'},z^{'},\theta^{'})-U_{L_\epsilon}(z^{'},\theta^{'}), \Big(\phi-\Hc_{\epsilon}\phi\Big)(t^{'},z^{'},\theta^{'})  \Big\}\nonumber\\
&\leq &\Min\Big\{ \phi(T,z,\theta)-U_{L_\epsilon}(z,\theta),\Big(\phi-\Hc_{\epsilon}\phi\Big)(T,z,\theta)\Big\}
\end{eqnarray}
and 
\begin{eqnarray}\label{defconsistent4}
& \Liminf_{\stackrel{(t^{'},z^{'},\theta^{'})\rightarrow(T,z,\theta)}{ (t^{'},z^{'},\theta^{'})\in [0,T)\times \Sc_{\epsilon}}}&
\Min\Big\{\phi(t^{'},z^{'},\theta^{'})-U_{L_\epsilon}(z^{'},\theta^{'}), \Big(\phi-\Hc_{\epsilon}\phi\Big)(t^{'},z^{'},\theta^{'})  \Big\}\nonumber\\
&\geq &\Min\Big\{ \Big(\phi(T,z,\theta)-U_{L_\epsilon}(z,\theta)),\Big(\phi-\Hc_{\epsilon}\phi\Big)(T,z,\theta) \Big\}
\end{eqnarray}
\end{Proposition}
\noindent {\bf Proof.}  The arguments are standard, and can be adapted  e.g. from  \cite{chaokssul02} or \cite{chefor08}. We sketch the proof, and   only show the inequality \reff{defconsistent1} since the other ones are derived similarly. Fix $t$ $\in$ $[0,T)$.  
Since the minimum of two upper-semicontinous (usc) functions  is also usc and using the caracterization of usc functions, we have
\beq\label{cons}
& &\Limsup_{\tiny{\stackrel{(h,t^{'},z^{'},\theta^{'})\rightarrow(0,t,z,\theta)}{ (t^{'},z^{'},\theta^{'})\in [0,T) \times  \Sc_{\epsilon}}}}
\Min\Big\{ \Big(\phi-\Hc_{\epsilon}\phi\Big)(t^{'},z^{'},\theta^{'}), \frac {\phi(t^{'},z^{'},\theta^{'})
- \E\Big[\phi(t^{'}+h , Z^{0,t^{'},z^{'}}_{t^{'}+h},\Theta^{0,t^{'},\theta^{'}}_{t^{'}+h} )\Big] }{h} \Big\}\nonumber\\
&\leq &\Limsup_{\stackrel{(h,t^{'},z^{'},\theta^{'})\rightarrow(0,t,z,\theta)}{ (t^{'},z^{'},\theta^{'})\in [0,T) \times  \Sc_{\epsilon}}}
\Min\Big\{\Limsup_{\stackrel{(h,t^{''},z^{''},\theta^{''})\rightarrow(0,t^{'},z^{'},\theta^{'})}{ (t^{''},z^{''},\theta^{''})\in [0,T) \times  \Sc_{\epsilon}}} 
\Big(\phi-\Hc_{\epsilon}\phi\Big)(t^{''},z^{''},\theta^{''}), \nonumber\\
& & \;\;\;\;\;
\Limsup_{\stackrel{(h,t^{''},z^{''},\theta^{''})\rightarrow(0,t^{'},z^{'},\theta^{'})}{ (t^{''},z^{''},\theta^{''})\in [0,T) \times  \Sc_{\epsilon}}} 
\frac {\phi(t^{''},z^{''},\theta^{''})
- \E\Big[\phi(t^{''}+h , Z^{0,t^{''},z^{''}}_{t^{''}+h},\Theta^{0,t^{''},\theta^{''}}_{t^{''}+h} )\Big] }{h} \Big\}\nonumber\\
&\leq& \Min\Big\{ \Limsup_{\stackrel{(h,t^{'},z^{'},\theta^{'})\rightarrow(0,t,z,\theta)}{ (t^{'},z^{'},\theta^{'})\in [0,T) \times  \Sc_{\epsilon}}}
\Big(\phi-\Hc_{\epsilon}\phi\Big)(t^{'},z^{'},\theta^{'}),\nonumber\\
& &\;\;\;\;\;  \Limsup_{\stackrel{(h,t^{'},z^{'},\theta^{'})\rightarrow(0,t,z,\theta)}{ (t^{'},z^{'},\theta^{'})\in [0,T) \times  \Sc_{\epsilon}}} 
\frac {\phi(t^{'},z^{'},\theta^{'})
- \E\Big[\phi(t^{'}+h , Z^{0,t^{'},z^{'}}_{t^{'}+h},
\Theta^{0,t^{'},\theta^{'}}_{t^{'}+h} )\Big] }{h} \Big\}  \nonumber \\
& \leq &   \Min\Big\{  \phi(t,z,\theta)-\Hc_{\epsilon}\phi(t,z,\theta) \nonumber \\
& &\;\;\;\;\;  \Limsup_{\stackrel{(h,t^{'},z^{'},\theta^{'})\rightarrow(0,t,z,\theta)}{ (t^{'},z^{'},\theta^{'})\in [0,T) \times  \Sc_{\epsilon}}} 
\frac {\phi(t^{'},z^{'},\theta^{'})
- \E\Big[\phi(t^{'}+h , Z^{0,t^{'},z^{'}}_{t^{'}+h},
\Theta^{0,t^{'},\theta^{'}}_{t^{'}+h} )\Big] }{h} \Big\}, 
\enq
where the last inequality follows from the continuity of $\phi$ and the lower semicontinuity of $\Hc_\eps$. Moreover,  by It\^o's formula applied to 
$\phi(s,Z_s^{0,t',z'},\Theta_s^{0,t',\theta'})$,   and standard arguments of localization to remove in expectation  the stochastic integral, we get 
\beqs \label{cons2membre}
\Limsup_{\stackrel{(h,t^{'},z^{'},\theta^{'})\rightarrow(0,t,z,\theta)}{ (t^{'},z^{'},\theta^{'})\in [0,T) \times  \Sc_{\epsilon}}} 
\frac {\phi(t^{'},z^{'},\theta^{'})
- \E\Big[\phi(t^{'}+h , Z^{0,t^{'},z^{'}}_{t^{'}+h},
\Theta^{0,t^{'},\theta^{'}}_{t^{'}+h} )\Big] }{h} 
&=&-\Big( \Dt{\phi} + \Lc \phi \Big)(t,z,\theta)
\enqs 
Substituting  into \reff{cons}, we obtain the desired inequality \reff{defconsistent1}.
 \ep

 \vspace{2mm}

Since the numerical scheme \reff{scheme1} is  monotone, stable and consistent, we can follow the viscosity solutions arguments as in \cite{barsou91} 
to prove the convergence of $v^h$ to $v_\eps$, by relying  on the PDE characterization of $v_\eps$ in Theorem \ref{theomainapprox} (2), 
and the strong  comparison principle for \reff{QVIvareps}-\reff{termeps} proven in  \cite{khapha09}.

\begin{Theorem} (Convergence) 
The solution $v^{h}$ of the numerical scheme \reff{scheme1} converges locally 
uniformly to $v_\eps$ on   $[0,T)\times \Sc_{\epsilon}$.
\end{Theorem} 
\noindent {\bf Proof.}
Let $\overline {v_{\epsilon}}$ and $\underline {v_{\epsilon}}$ be defined on $[0,T]\times\bar\Sc_\eps$ by
\beqs
\overline {v_{\epsilon}}(t,z,\theta) &= & \Limsup_{\tiny{\stackrel{(h,t^{'},z^{'},\theta^{'})\rightarrow(0,t,z,\theta)}
{ (t^{'},z^{'},\theta^{'})\in [0,T)\times \Sc_{\epsilon}}}}  
v^h (t^{'},z^{'},\theta^{'}) \\ 
\underline {v_{\epsilon}}(t,z,\theta) &= & \Liminf_{\tiny{\stackrel{(h,t^{'},z^{'},\theta^{'})\rightarrow(0,t,z,\theta)}
{ (t^{'},z^{'},\theta^{'})\in [0,T)\times \Sc_{\epsilon}}}} 
v^h (t^{'},z^{'},\theta^{'})
\enqs 
We first  see that $\overline {v_{\epsilon}}$ and $\underline {v_{\epsilon}}$ are respectively viscosity subsolution and 
supersolution of \reff{QVIvareps}-\reff{termeps}.  
These viscosity properties follow indeed, by standard arguments as in \cite{barsou91} (see also \cite{chaokssul02} or \cite{chefor08} for impulse control problems), from the monotonicity, stability and consistency properties.  Details can be obtained upon request to the authors. 
Moreover,  from \reff{defvh}, we have the inequality:  $U(0)$ $\leq$ $v^h$ $\leq$ $v_\eps$, which implies by \reff{bounveps}: 
\begin{eqnarray}\label{inecomp1}
\Liminf_{\stackrel{(t^{'},z^{'},\theta^{'})\rightarrow(t,z,\theta)}
{ (t^{'},z^{'},\theta^{'})\in [0,T)\times \Sc_{\epsilon}}} 
\underline{v_{\epsilon}}(t^{'},z^{'},\theta^{'}) &= & U(0) \; = \;   \overline{v_{\epsilon}}(t,z,\theta),  \;\,\,\forall\, (t,z,\theta)\in [0,T]\times D_0
\end{eqnarray}
Thus, by using the strong comparison principle for \reff{QVIvareps}-\reff{termeps} stated in Theorem 5.2 \cite{khapha09}, we deduce that 
$\overline {v_{\epsilon}}\leq \underline  {v_{\epsilon}}$  on $[0,T]\times \Sc_{\epsilon}$ and so 
$\overline {v_{\epsilon}}= \underline  {v_{\epsilon}}=v_{\epsilon}$ on  $[0,T]\times \Sc_{\epsilon}$. This proves the required convergence result. 
\ep

\section{Numerical Algorithm}

\setcounter{equation}{0} 
\setcounter{Assumption}{0} 
\setcounter{Example}{0} 
\setcounter{Theorem}{0} 
\setcounter{Proposition}{0}
\setcounter{Corollary}{0} 
\setcounter{Lemma}{0} 
\setcounter{Definition}{0} 
\setcounter{Remark}{0}

Let us consider a time  step $h$ $=$ $T/m$, $m$ $\in$ $\N\setminus\{0\}$, and denote by 
$\T_m$ $=$ $\{t_i=ih, i=0,\ldots,m\}$ the regular grid over the interval $[0,T]$. 
We recall from the previous section  that the time discretization of step $h$  
for  the QVI \reff{QVIvareps}-\reff{termeps} leads to the convergent 
explicit backward scheme: 
\beq
v^h(t_m,z,\theta) &=& \left\{ \begin{array}{ll}
				  	U_{L_\eps}(z,\theta) &\;  \mbox{ if } \; \theta = 0 \\
				\max\Big[  U_{L_\epsilon}(z,\theta) \;  ,  & \\   
		\;\;\;\;\;\;\;\;\;\;	\Sup_{\tiny{e\in\Cc_\eps(z,\theta)}}  v^h(t_m,\Gamma_\eps(z,\theta,e),0) \Big],  & \; \mbox{ if } \; \theta > 0, 
				\end{array}
				\right.  \label{vhterm3} \\								 
v^h(t_i,z,\theta) &=&   \left\{ \begin{array}{ll} 
					\E\big[v^h(t_{i+1} ,Z^{0,t_i,z}_{t_{i+1}},\theta + h) \big]  & \; \mbox{ if } \;\theta = 0 \\
					\max\Big[ \E\big[v^h(t_{i+1} ,Z^{0,t_i,z}_{t_{i+1}},\theta + h) \big] \; , &    \\
             \;\;\;\;\;\;\;\;\;   \Sup_{e\in\Cc_\eps(z,\theta)}  v^h(t_{i} , \Gamma_\eps(z,\theta,e), 0 ) \big]   \Big], & \; \mbox{ if } \; \theta > 0 
             				\end{array}
				\right.  \label{vht3}
\enq
for $i$ $=$ $0,\ldots,m-1$,   $(z=(x,y,p),\theta)$ $\in$ $\bar\Sc_\eps$.   Recall that the variable $\theta$ 
represents the time lag between the current time $t$ and the last trade. Thus, it  suffices to consider  at each time step $t_i$ of 
$\T_m$, a discretization for $\theta$ valued in the time grid
 \beqs
 \T_{i,m} &=& \big\{ \theta_j=jh, \;\; j = 0,\ldots,i \}, \;\;\; i = 0,\ldots,m. 
 \enqs
On the other hand, the above scheme involves nonlocal terms in  the variable $z$ for the solution $v^h$ in   
relation with the supremum over $e\in\Cc_\eps(z,\theta)$ and the expectations in \reff{vhterm3}-\reff{vht3}, and thus the practical implementation requires a  discretization  for the state variable  $z$, together with an interpolation. 
For any $\theta_j$ $\in$ $\T_{i,m}$,  let us denote by 
\beqs
\Zc^j &=& \big\{ z =(x,y,p) \in \R\times\R_+\times \R_+: (z,\theta_j) \in \bar\Sc_\eps \big\}. 
\enqs
For the discretization of the state variable $z$ $\in$ $\Zc^j$, and  since $\bar\Sc_\eps$ is unbounded, we first  
localize  the domain by  setting $\Zc^j_{loc}$ $=$ $\Zc^j$ $\cap$ $([x_{min},x_{max}]\times [y_{min},y_{max}]\times [p_{min},p_{max}])$, where 
$x_{min}$ $<$ $x_{max}$ in $\R$, $0$ $\leq$ $y_{min}$  $<$ $y_{max}$,  $0$ $\leq$ $p_{min}$ $<$ $p_{max}$ are fixed constants, and 
then define the regular grid: 
\beqs
\Z_n^j &=& \big\{ z =(x,y,p) \in \X_n \times \Y_n \times \P_n: (z,\theta_j) \in \bar\Sc_\eps\big\}. 
\enqs
 where $\X_n$ is the uniform grid on $[x_{min},x_{max}]$ of step $\frac{x_{max}-x_{min}}{n}$, and similarly for $\Y_n$, $\P_n$.
 
 \vspace{2mm}
 
 \noindent {\bf Optimal quantization  method.} 
 Let us now describe the numerical procedure for com\-puting the expectations arising in \reff{vht3}. 
 Recalling that $Z^{0,t,z}$ $=$ $(x,y,P^{t,p})$,  this involves only the expectation with respect to the price process, assumed here to follow 
 a Black-Scholes model \reff{dynP0}. We shall then use an optimal quantization for the standard normal random variable $U$, which consists in approximating  the distribution of $U$ by the discrete law  of a random variable $\hat U$ of support  $(u_k)_{1\leq k\leq N}$ $\in$ $\R^N$, and  defined as the projection of $U$ on the grid $(u_k)_{1\leq k\leq N}$  according to the  closest neighbour.  
 The grid $(u_k)_{1\leq k\leq N}$  is optimized in order to minimize the  distorsion error, i.e. the quadratic  norm between $U$ and $\hat U$.  This optimal grid and the associated 
 weights  $(\pi_k)_{1\leq k\leq N}$  are downloaded from the website:  http://87.106.220.249/n01. We refer to the survey article 
 \cite{pagphapri04} for more details on the theoretical and computational aspects of optimal quantization methods.  From \reff{vht3}, we have to compute at any time  step $t_i$ $\in$ $\T_m$, and for any $\theta_j$ $\in$ $\T_{i,m}$, $z$ $=$ $(x,y,p)$ $\in$ $\Z_n^j$, expectations in the form: 
 \beqs
 \Ec^h(t_i,z,\theta_j) &:=&  \E\big[v^h(t_{i}+h ,Z^{0,t_i,z}_{t_{i}+h},\theta_j + h) \big] \\
 &=&  \E\big[v^h(t_{i}+h ,x,y,p \exp\big((b - \frac{\sigma^2}{2})h + \sigma \sqrt{h} U \big),\theta_j + h) \big], 
 \enqs 
 that we approximate by
 \beq \label{expap}
 \Ec^h(t_i,z,\theta_j) & \simeq &  \frac{1}{N} \sum_{k=1}^{N} \pi_k \; 
 v^h(t_i+h , x , y ,p \exp\big((b - \frac{\sigma^2}{2})h + \sigma \sqrt{h} u_k \big),\theta_j + h ). 
 \enq
 
 \vspace{1mm}
 
 \noindent {\bf Interpolation procedure.} 
 Notice that  for implementing recursively in \reff{vht3} this quantization  method, we  need  to compute  $z$ $\rightarrow$ 
 $v^h(t_i,z,\theta_j)$ on $\Zc^j$ given  the known values  of $v^h(t_i,z,\theta_j)$ on $\Z_n^j$. This approximation is achieved as follows: 
\begin{itemize}
\item if  $z$ $\in$ $\Zc_{loc}^j$,  then we use a linear interpolation of $v^h(t_i,z,\theta_j)$ with respect to the closest neighbours in $\Z_n^j$. 
\item if $z$ $=$ $(x,y,p)$   $\notin$ $\Zc_{loc}^j$, we use the growth condition satisfied by the value function: 
\beqs
v^h(t_i,z,\theta_j) &\simeq& v^h(t_i,\hat{z},\theta_j) \frac{(x +  yp)^\gamma}{(\hat{x}+ \hat y\hat p)^\gamma}
\enqs
where $\hat z$ $=$ $(\hat x,\hat y,\hat p)$  is the projection of $z$ (according to the closest neighbour) on the grid $\Z_n^j$. 
\end{itemize}

 \vspace{2mm}

\noindent {\bf Algorithm description.}  In summary, our numerical scheme provides  an algorithm for computing approximations $v^h$ of the value function, and $\zeta^h$ of the optimal trading strategy at each time step $t_i$ $\in$ $\T_m$, and each point $(z,\theta)$ of the grid 
$(\Z_n\times\T_{i,m})$ $\cap$ $\bar\Sc_\eps$. The parameters in the algorithm are:
 
 \vspace{1mm}

\noindent  - $T$ the maturity

\noindent - $b$ and $\sigma$ the Black and Scholes parameters of the stock price

\noindent - $\lambda$ the impact parameter, $\beta$ the impact exponent in the market impact function \reff{Qexp}
 
\noindent - $\kappa_a$, $\kappa_b$ the spread parameters in percent, $\eps$ the transactions costs fee

\noindent -  We take by default a  CRRA utility function: $U(x)$ $=$ $x^\gamma$

\noindent -  $x_{min},x_{max}$ $\in$ $\R$, $0\leq y_{min} <y_{max}$, $0\leq p_{min}<p_{max}$, the boundaries of the localized domain

\noindent -  $m$  number of steps in time discretization, $n$ the number of steps in space discretization

\noindent -  $N$ number of points for optimal quantization of the normal law, $M$  number of points used in the static supremum  in $e$

\vspace{2mm}

The algorithm is described explicitly in backward induction as follows:

\vspace{1mm}

\noindent  $\trib$ {\it Initialization step at time $t_m$ $=$ $T$:}
\begin{itemize}
\item {\it (s:0)} For $j$ $=$ $0$,  set $v^h(t_m,z,0)$ $=$  $U_{L_\eps}(z,0)$, $\zeta^h(t_m,z,0)$ $=$ $0$ on $\Z_n^0$, 
and interpolate $v^h(t_m,z,0)$ on 
$\Zc^0$.
\item {\it (s:j)} For $j$ $=$ $1,\ldots,m$,  
 	\begin{itemize}
	\item   for $z$ $\in$ $\Z_n^j$, compute $v$ $:=$ $\Sup_{e\in\Cc_\eps(z,\theta_j)} U_{L_\eps}(\Gamma_\eps(z,\theta_j,e),0)$ and 
	denote by $\hat e$ the argument maximum:
	\item   if $v$ $>$ $U_{L_\epsilon}(z,\theta_j)$, then set $v^h(t_m,z,\theta_j)$ $=$ $v$ and $\zeta^h(t_m,z,\theta_j)$ $=$ $\hat e$,  
	\item   else set $v^h(t_m,z,\theta_j)$ $=$ $U_{L_\eps}(z,\theta_j)$, and $\zeta^h(t_m,z,\theta_j)$ $=$ $0$.
	\item   Interpolate  $z$ $\rightarrow$ $v^h(t_m,z,\theta_j)$ on $\Zc^j$. 
	\end{itemize}
\end{itemize}

\noindent  $\trib$ {\it From time step $t_{i+1}$ to $t_i$, $i$ $=$ $m-1,\ldots,0$:}
\begin{itemize}
\item {\it (s:0)} For $j$ $=$ $0$, compute $\Ec^h(t_i,z,0)$ from \reff{expap} and {\it (s:1)} of time step $t_{i+1}$, and set  $v^h(t_i,z,0)$ $=$ 
$\Ec^h(t_i,z,0)$,  $\zeta^h(t_i,z,0)$ $=$ $0$ on 
$\Z_n^0$; interpolate $v^h(t_i,z,0)$ on $\Zc^0$.
\item {\it (s:j)} For $j$ $=$ $1,\ldots,i$, 
	\begin{itemize}
	\item for  $z$ $\in$ $\Z_n^j$, compute  $\Ec^h(t_i,z,\theta_j)$ from \reff{expap} and {\it (s:j+1)} of time step $t_{i+1}$,  $v$ $:=$ 
	$\Sup_{e\in\Cc_\eps(z,\theta_j)} v^h(t_i,\Gamma_\eps(z,\theta_j,e),0)$ from {\it (s:0)}, and denote by  $\hat e$ the argument maximum:
	\item if  $v$ $>$  $\Ec^h(t_i,z,\theta_j)$, then set $ v^h(t_i,z,\theta_j)$ $=$ $v$, and  $\zeta^h(t_i,z,\theta_j)$ $=$ $\hat e$, 
	\item else set $v^h(t_i,z,\theta_j)$ $=$  $\Ec^h(t_i,z,\theta_j)$, and $\zeta^h(t_i,z,\theta_j)$ $=$ $0$.
	\item Interpolate  $z$ $\rightarrow$ $v^h(t_i,z,\theta_j)$ on $\Zc^j$. 
	\end{itemize}
\end{itemize}

\section{Numerical Results}

\setcounter{equation}{0} 
\setcounter{Assumption}{0} 
\setcounter{Example}{0} 
\setcounter{Theorem}{0} 
\setcounter{Proposition}{0}
\setcounter{Corollary}{0} 
\setcounter{Lemma}{0} 
\setcounter{Definition}{0} 
\setcounter{Remark}{0}


\subsection{Procedure}

For each of the numerical tests, we used the same procedure consisting in the following steps:


\vspace{1mm}

\noindent (1) Set the parameters according to the parameter table  described in the first subsection of each test

\vspace{1mm}

\noindent (2) Compute and save the grids representing value function and optimal policy according to the optimal liquidation algorithm

\vspace{1mm}

\noindent (3) Generate $Q$ paths for the stock price process following a geometrical Brownian motion: we choose parameters $b$ and $\sigma$  
that allows us to observe several empirical facts on the performance and the behavior of optimal liquidation strategy. These parameters can also be estimated from historical observations on real data  by standard statistical methods.  
 
\vspace{1mm}

\noindent (4) Consider the portfolio made of $X_0$ dollars and $Y_0$ shares of risky asset

\vspace{1mm}

\noindent (5) For each price path realization, update the portfolio along time and price path accordingly to the policy computed in the second step

\vspace{1mm}

\noindent (6) Save each optimal liquidation realization

\vspace{1mm}

\noindent (7) Compute statistics

\vspace{2mm}

 In the sequel, we shall use  the following quantities  as descriptive statistics:

\begin{itemize}
\item The  performance of the $i$-th realization of the optimal strategy is defined by
\beqs
L_{opt}^{(i)} &=&  \dfrac{L_\epsilon(Z_T^{(i),\alpha^{opt}},\Theta_T^{(i),\alpha^{opt}})}{X_0 + Y_0 P_0}
\enqs
\end{itemize}
where $ (Z_T^{(i),\alpha^{opt}},\Theta_T^{(i),\alpha^{opt}}) $ is the state process, starting at date $0$ at $(X_0,Y_0,P_0,0)$, 
evolving under the $i$-th price realization and the optimal control $\alpha^{opt}$. This quantity can be interpreted as the ratio between the cash obtained from the optimal liquidation strategy and the ideal Merton liquidation. We define in the same way the quantities $L_{naive}^{(i)}$ and $L_{uniform}^{(i)}$ respectively associated with the controls $\alpha^{naive}$ and $\alpha^{uniform}$ of the naive and uniform strategy, refereed  as benchmark strategies. 
 Recall that 
the naive strategy consists in liquidating the whole portfolio in one block at the last date, and the uniform strategy 
consists in liquidating the same quantity of asset at each predefined date until the last date. Notice that 
the score $1$ corresponds to  the strategy, which  consists in liquidating the whole portfolio immediately in an ideal Merton market. 


 \vspace{1mm}

When denoting by $Q$ the number of paths of our simulation, we define:
\begin{itemize}
\item The mean utility $\displaystyle \hat{V}_. \; = \;  \frac{1}{Q} \sum_{i=1}^Q  U(L_.^{(i)})$
\item The mean performance $\displaystyle \hat{L}_. \;=\; \frac{1}{Q} \sum_{i=1}^Q  L_.^{(i)}$
\item The standard deviation of the strategy $\displaystyle \hat{\sigma}_. \; = \;  \sqrt{\frac{1}{Q} \sum_{i=1}^Q  (L_.^{(i)})^2 - \hat{L_.}^2}$
\end{itemize}
Here the dot $.$ stands for $opt$, $naive$ or $uniform$. We will also compute the third and fourth standardized moments for the series $(L_.^{(i)})_i$.


\subsection{Test 1: A toy example}
The goal of this test is to show the main characteristics of our results. We choose a set of parameters that is unrealistic but that has the advantage of emphasizing the typical behavior of the optimal liquidation strategy. 
\paragraph{Parameters} 
We choose the set of parameters shown in table \ref{T1PARAM}. 

\begin{table}[h!]
\begin{center}
\begin{tabular}{|ll|ll|}
\hline Parameter & Value & Parameter & Value \\ 
\hline  Maturity & 1 year & $X_0$ & 2000 \\ 
$\lambda$ &		5.00E-07& $Y_0$ & 2500 \\
$\beta$ &			0.5 & $P_0$ & 5.0 \\
$\gamma$ &			0.5 & $x_{min}$ & -30000  \\
$\kappa_A$ &		1.01 & $x_{max}$ & 80000  \\
$\kappa_B$&			0.99 & $ y_{min}$ & 0  \\
$\epsilon$&			0.001 & $ y_{max }$& 5000   \\
$b$		&		0.1 & $p_{min}$& 0  \\
$\sigma$	&		0.5 & $p_{max}$ & 20  \\
&  & $m$  & 40\\
&  & $n$  & 20\\
&  & $N$  & 100\\
&  & $Q$  & $10^5$\\
\hline 
\end{tabular}
\end{center} 
\caption{Test 1: parameters}
\label{T1PARAM}
\end{table}
\paragraph{Execution statistics}
The results were computed using Intel$^\circledR$ Core 2 Duo at 2.93Ghz CPU with 2.98 Go of RAM. Statistics are shown in table \ref{T1EXECSTATS}.
\begin{table}[h!]
\begin{center}
\begin{tabular}{ll}
\hline Quantity & Evaluation \\
\hline Time Elapsed for grid computation in seconds &	7520\\
Number Of Available Processors &	2\\
Estimated Memory Used (Upper bound) &	953MB\\
Time Elapsed for statistics Computation in seconds	& 21\\
\hline 
\end{tabular}
\end{center}
\caption{Test 1: Execution statistics}
\label{T1EXECSTATS}
\end{table}

\begin{figure}[h!]
\centering
\includegraphics[width=0.9\textwidth]{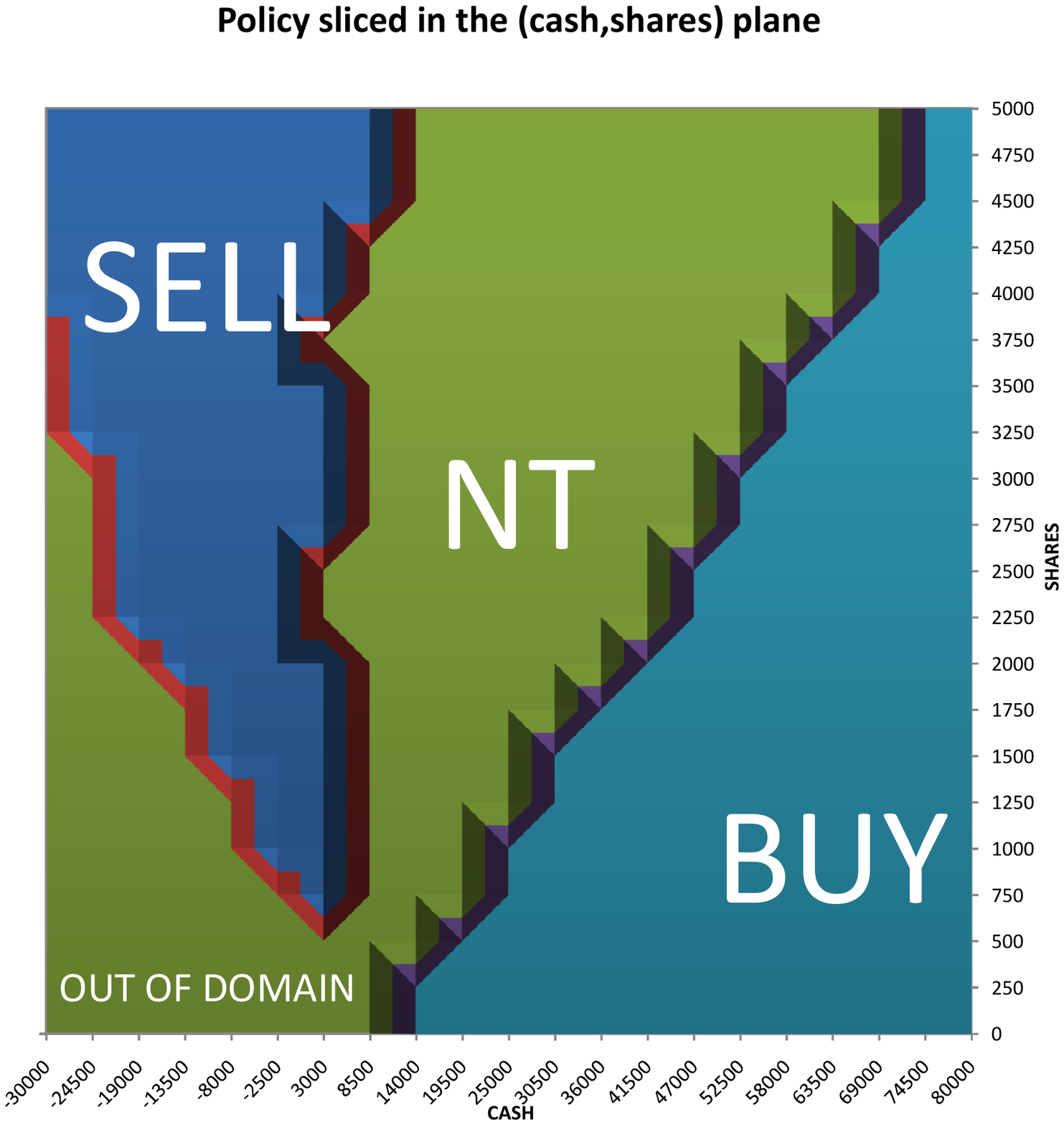} 
\caption{Test 1: Typical shape of the policy sliced in XY near date 0}
\label{policyXYstart}
\end{figure}

\paragraph{Shape of policy}  In this paragraph we plotted the shape of the policy sliced in the plane $(x,y)$, i.e. the (cash, shares) plane, for a fixed $(t,\theta,p)$ (figure \ref{policyXYstart}). The color of the map at $(x_0,y_0)$ on the graph represents the action one has to take when reaching the state $(t,\theta,x_0,y_0,p)$. We can see three zones: a buy zone (denoted BUY on the graph), a sell zone (denoted SELL on the graph) and a no trade zone (denoted NT on the graph). Note that the bottom left zone on the graph is outside the domain  $\bar{\mathcal{S}}$. These results have the intuitive   financial interpretation: when $x$ is big and $y$ is small, the investor has enough  cash to buy shares of the risky asset and tries to profit from an increased exposure. When $y$ is large and $x$ is small, the investor has to reduce exposure to match the terminal liquidation constraint.

\begin{figure}[h!]
\centering
\includegraphics[width=0.9\textwidth]{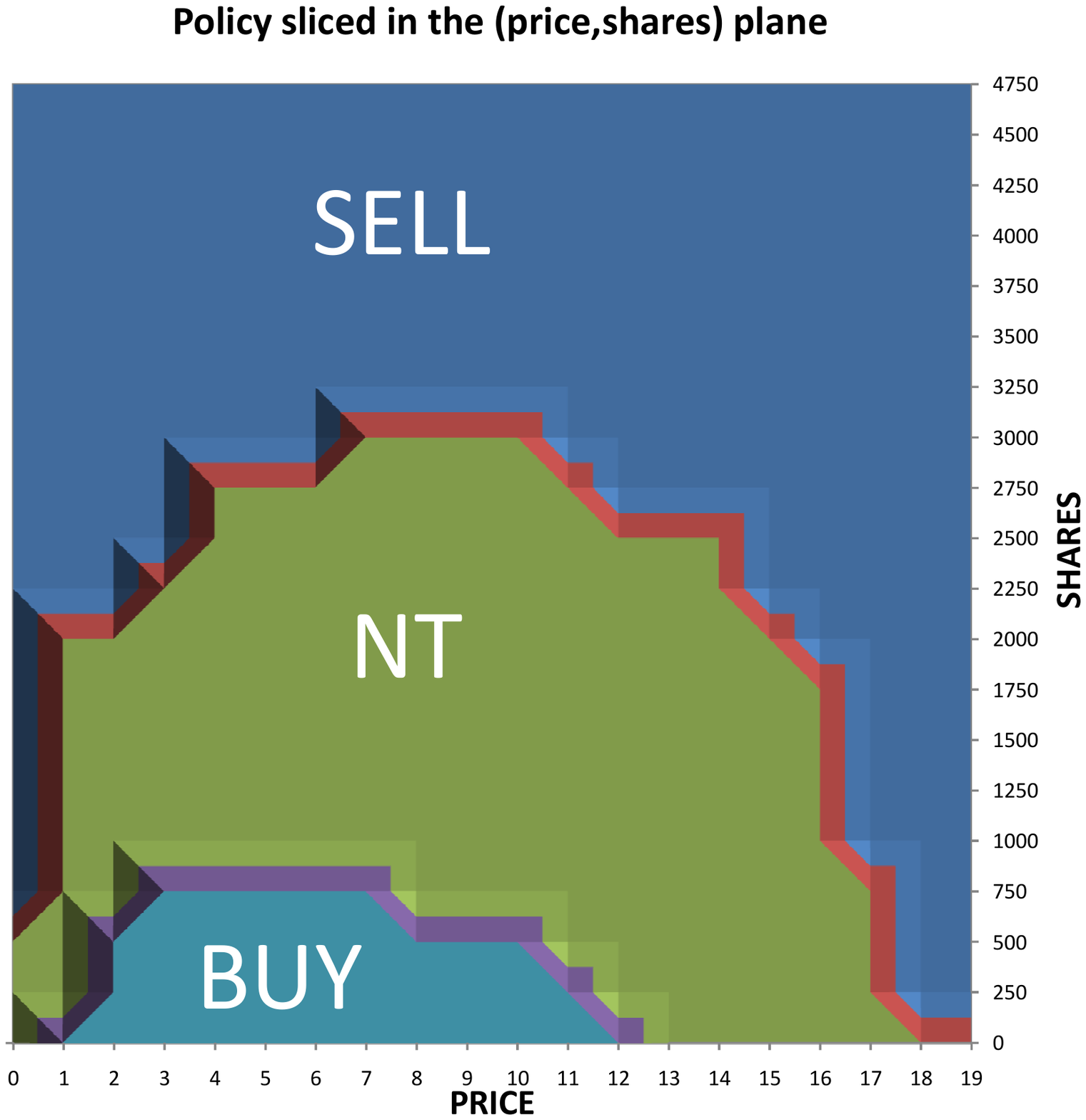} 
\caption{Test 1: Typical shape of the policy sliced in YP}
\label{policyYP}
\end{figure}

We also plotted the shape of the policy sliced in the plane $(y,p)$, i.e. the (shares,price) plane, for a fixed $(t,\theta,x)$ (figure \ref{policyYP}). As before, the color of the map at $(y_0,p_0)$ on the graph represents the action one has to take when reaching the state $(t,\theta,x,y_0,p_0)$. Again, we can distinguish the three zones: buy, sell and no trade. 

\begin{figure}[h!]
\centering
\includegraphics[width=0.9\textwidth]{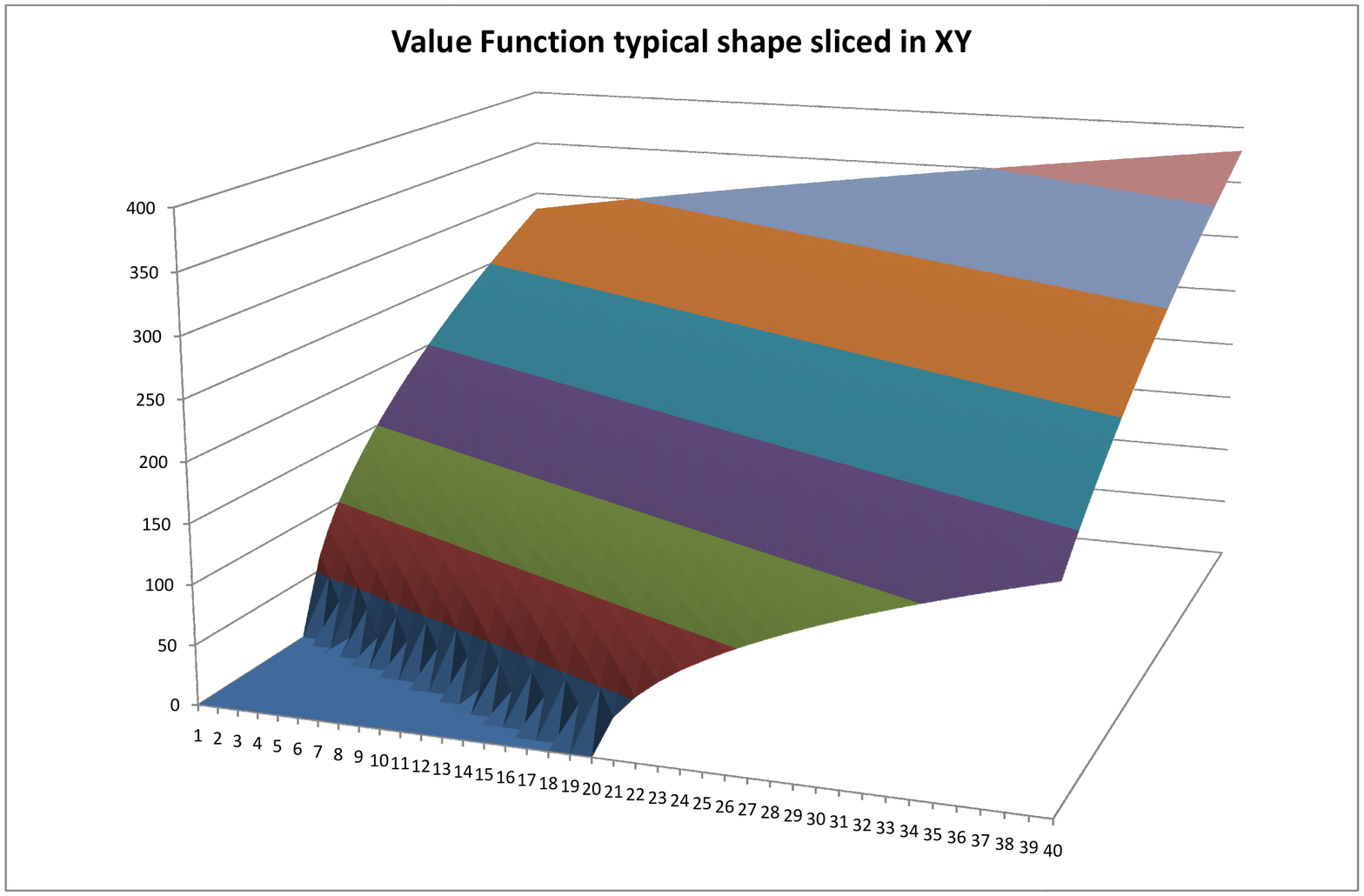} 
\caption{Test 1: Typical shape of the value function sliced in XY}
\label{valueXY}
\end{figure}

\begin{figure}[h!]
\centering
\includegraphics[width=0.9\textwidth]{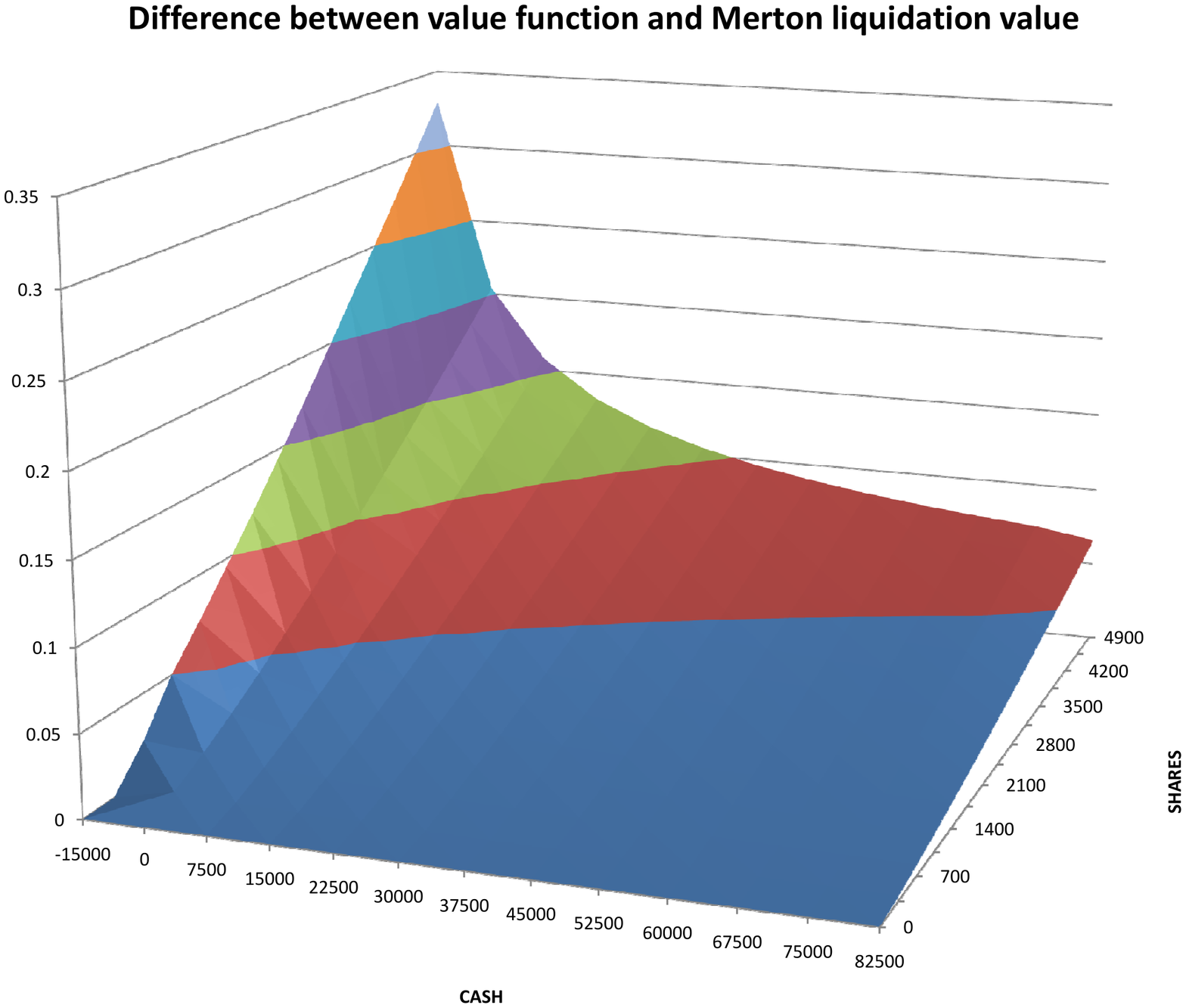} 
\caption{Test 1: Difference between value function and Merton theoretical bound}
\label{diffmerton}
\end{figure}

\paragraph{Shape of value function}
Figure \ref{valueXY} shows the value function sliced in the $(x,y)$ plane. This figure is a  typical pattern of the value function. 
Recall from Proposition 3.1 in \cite{khapha09})  the following Merton theoretical bound for the value function: 
\beqs 
v(t,z,\theta) &\leq&  v_M(t,x,y,p) \; = \; e^{\rho (T-t) } (x+yp)^\gamma, \;\;\; \mbox{ with }  \; 
\rho \; = \; \dfrac{\gamma}{1-\gamma} \dfrac{b^2}{2\sigma^2}.
\enqs
In the figure \ref{diffmerton} we plotted the difference between the value function and this theoretical bound. We observe that this difference is increasing with the number of shares, and decreasing with the cash. This result is interpreted as follows: the price impact increases with the number of shares, but this can be reduced by the liquidation strategy whose efficiency is greater if the investor can sustain bigger cash variations.

\subsection{Test 2: Short term liquidation}
The goal of this test is to show the behavior of the algorithm on a realistic set of parameters and real data. We used Reuters$\texttrademark$ data fed by OneTick$\texttrademark$ TimeSeries Database. We used the spot prices (Best Bid and Best Ask) for the week starting 04/19/2010 on BNP.PA. We computed mid-price that is the middle between best bid and best ask price. We choose the impact parameter $\lambda$ in order to penalize by approximately 1\% the immediate liquidation of the whole portfolio compared to Merton liquidation.  In other words,  we take $\lambda$ so that: 
$\lambda \vert \dfrac{Y_0}{T}\vert^\beta \simeq 0.01$.

\paragraph{Parameters}
We computed the strategy with parameters shown in table \ref{T2PARAM}.

\begin{figure}[h!]
\centering
\includegraphics[width=0.9\textwidth]{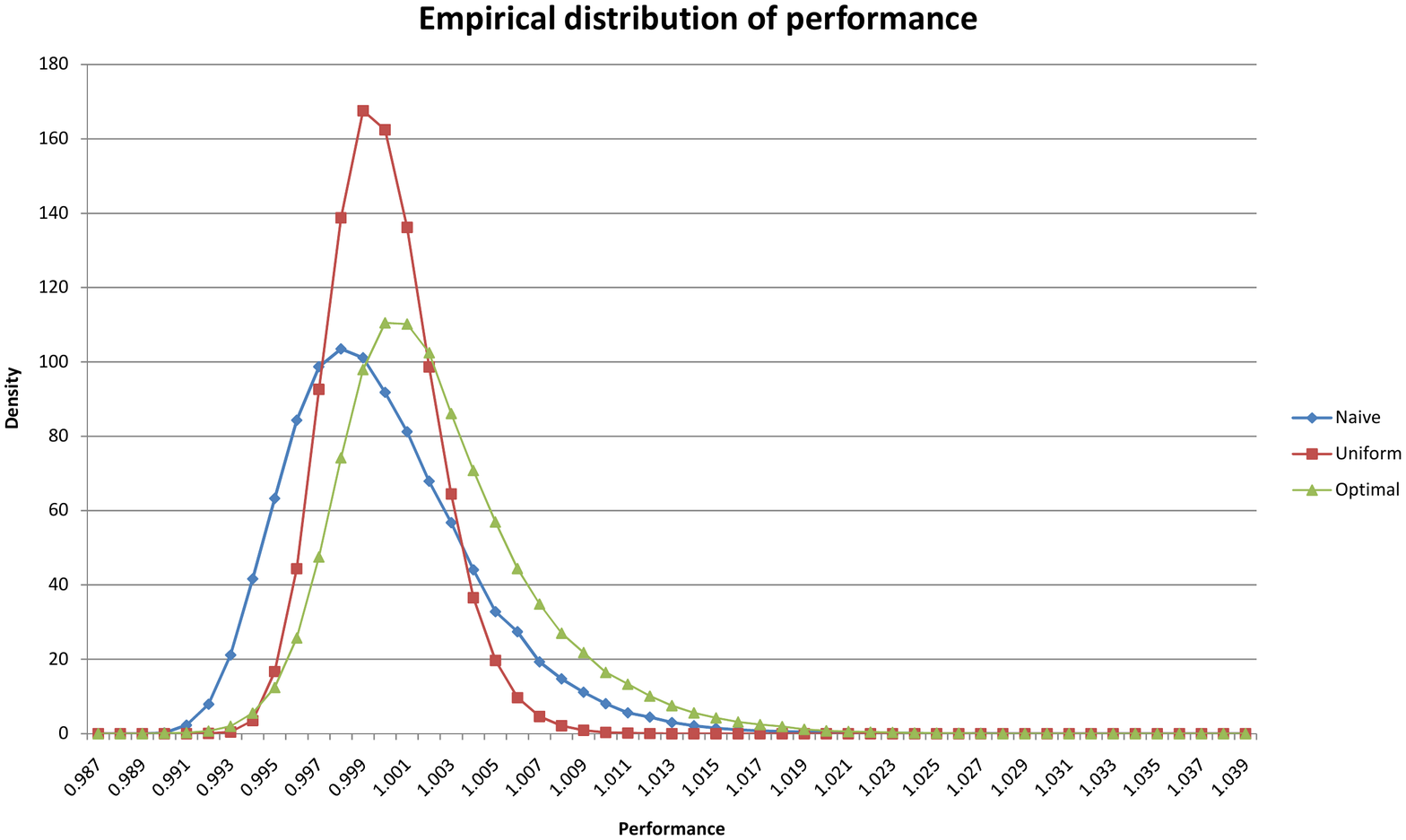} 
\caption{Test 2: Strategy empirical distribution }
\label{T2empdistrib}
\end{figure}

\begin{table}[h!]
\begin{center}
\begin{tabular}{|ll|ll|}
\hline Parameter & Value & Parameter & Value\\
\hline  Maturity & 1 Day & $X_0$ & 20000 \\ 
$\lambda$ &		5.00E-04 & $Y_0$ & 2500\\
$\beta$ &			0.2 & $P_0$ & 52.0 \\
$\gamma$ &			0.5 & $x_{min}$ & -30000 \\
$\kappa_A$ &		1.0001 & $x_{max}$ & 200000 \\
$\kappa_B$&			0.9999 & $ y_{min}$ & 0 \\
$\epsilon$&			0.001 & $ y_{max }$& 5000 \\
$b$		&		0.005 & $p_{min}$& 50.0 \\
$\sigma$	&		0.25 & $p_{max}$ & 54.0 \\
&  & $m$  & 30\\
&  & $n$  & 40\\
&  & $N$  & 100\\
&  & $Q$  & $10^5$\\
\hline 
\end{tabular}
\end{center}
\caption{Test 2: Parameters}
\label{T2PARAM}
\end{table}
\paragraph{Execution statistics}
We obtained the results using Intel$^\circledR$ Core 2 Duo at 2.93Ghz CPU with 2.98 Go of RAM, the computations statistics are gathered in table \ref{T2EXECSTATS}.
\begin{table}[h!]
\begin{center}
\begin{tabular}{ll}
\hline Quantity & Evaluation \\
\hline Time Elapsed for grid computation in seconds &	8123\\
Number Of Available Processors &	2\\
Estimated Memory Used (Upper bound) &	573MB\\
\hline 
\end{tabular}
\end{center}
\caption{Test 2: Execution statistics}
\label{T2EXECSTATS}
\end{table}
\paragraph{Performance Analysis}
We computed the mean utility and the first four moments of the optimal strategy and the two benchmark strategies in table \ref{T2PERF} and plotted the empirical distribution of performance in figure \ref{T2empdistrib}. It is remarkable that the optimal strategy gives an empirical performance that is above the immediate liquidation at date 0 in the Merton ideal market. This is due to the fact that the optimal strategy has an opportunistic behavior, as the decisions 
are based on the price level, and so  profit from the `detection" of some favorable price conditions. Indeed, an optimal trading strategy is embedded with the optimal liquidation: in this example, this feature not only compensates the trading costs, but also provides an extra performance compared to an ideal immediate liquidation at date 0. Still, the Merton case is a theoretical upper bound in the following sense: the optimal value function with trading costs is below the optimal value function without trading costs, recall the figure \ref{diffmerton}. As expected, the empirical distribution is between the distributions of the two other benchmark strategies. We also notice that the optimal strategy outperforms the two others by approximatively 0.25\% in utility and in performance. 

\begin{table}[h!]
\begin{center}
\begin{tabular}{|l|c|c|c|c|c|}
\hline Strategy & Utility $\hat{V}$ & Mean $\hat{L}$ & Standard Dev. & Skewness & Kurtosis \\ 
\hline Naive &  0.99993& 0.99986 & 0.00429 & 0.94584 & 4.68592 \\ 
\hline Uniform & 0.99994 &  0.99988 &  0.00240 & 0.42788  & 3.34397  \\ 
\hline Optimal & 1.00116 & 1.00233 & 0.00436 & 1.03892 & 4.89161 \\ 
\hline 
\end{tabular}
\end{center}
\caption{Test 2: Utility and first four moments for the optimal strategy and the two benchmark strategies}
\label{T2PERF}
\end{table}

We also computed other statistics in table \ref{otherstats}. 

\begin{table}[h!]
\begin{center}
\begin{tabular}{lll}
\hline Quantity & Formula & Value \\
\hline Winning percentage &	$\displaystyle \frac{1}{Q}\sum_{i=1}^Q 1_{\lbrace L_{opt}^{(i)} > \max(L_{naive}^{(i)},L_{uniform}^{(i)})\rbrace}$ &	58.8\%\\
Relative Optimal Utility & $ \dfrac{\hat{V}_{opt}-\max(\hat{V}_{naive},\hat{V}_{uniform})}{\hat{V}_{opt}}$&			0.00238\\
Relative Optimal Performance & $ \dfrac{\hat{L}_{opt}-\max(\hat{L}_{naive},\hat{L}_{uniform})}{\hat{L}_{opt}}$ &			0.00244\\
Utility Sharpe Ratio & $ \dfrac{\hat{V}_{opt}-\max(\hat{V}_{naive},\hat{V}_{uniform})}{\hat{\sigma}_{opt}}$&		0.28017\\
Performance Sharpe Ratio & $ \dfrac{\hat{L}_{opt}-\max(\hat{L}_{naive},\hat{L}_{uniform})}{\hat{\sigma}_{opt}}$ &			0.56140\\
VaR 95\% Naive Strategy 	&	$\displaystyle\sup\left\lbrace x \mid \frac{1}{Q}\sum_{i=1}^Q 1_{\lbrace L_{naive}^{(i)}>x \rbrace} \geq 0.95\right\rbrace$ &		0.994\\
VaR 95\% Uniform Strategy	&	$\displaystyle\sup\left\lbrace x \mid \frac{1}{Q}\sum_{i=1}^Q 1_{\lbrace L_{uniform}^{(i)}>x \rbrace} \geq 0.95\right\rbrace$&		0.996\\
VaR 95\% Optimal Strategy	&	$\displaystyle\sup\left\lbrace x \mid \frac{1}{Q}\sum_{i=1}^Q 1_{\lbrace L_{opt}^{(i)}>x \rbrace} \geq 0.95\right\rbrace$&		0.997 \\
VaR 90\% Naive Strategy 	&	$\displaystyle\sup\left\lbrace x \mid \frac{1}{Q}\sum_{i=1}^Q 1_{\lbrace L_{naive}^{(i)}>x \rbrace} \geq 0.90\right\rbrace$&		0.995\\
VaR 90\% Uniform Strategy	&	$\displaystyle\sup\left\lbrace x \mid \frac{1}{Q}\sum_{i=1}^Q 1_{\lbrace L_{uniform}^{(i)}>x \rbrace} \geq 0.90\right\rbrace$&		0.997\\
VaR 90\% Optimal Strategy	&	$\displaystyle\sup\left\lbrace x \mid \frac{1}{Q}\sum_{i=1}^Q 1_{\lbrace L_{opt}^{(i)}>x \rbrace} \geq 0.90\right\rbrace$&		0.998 \\
\hline 
\end{tabular}
\end{center}
\caption{Test 2: Other statistics on performance of optimal strategy}
\label{otherstats}
\end{table}


\paragraph{Behavior Analysis}

\begin{figure}[h!]
\centering
\includegraphics[width=0.9\textwidth]{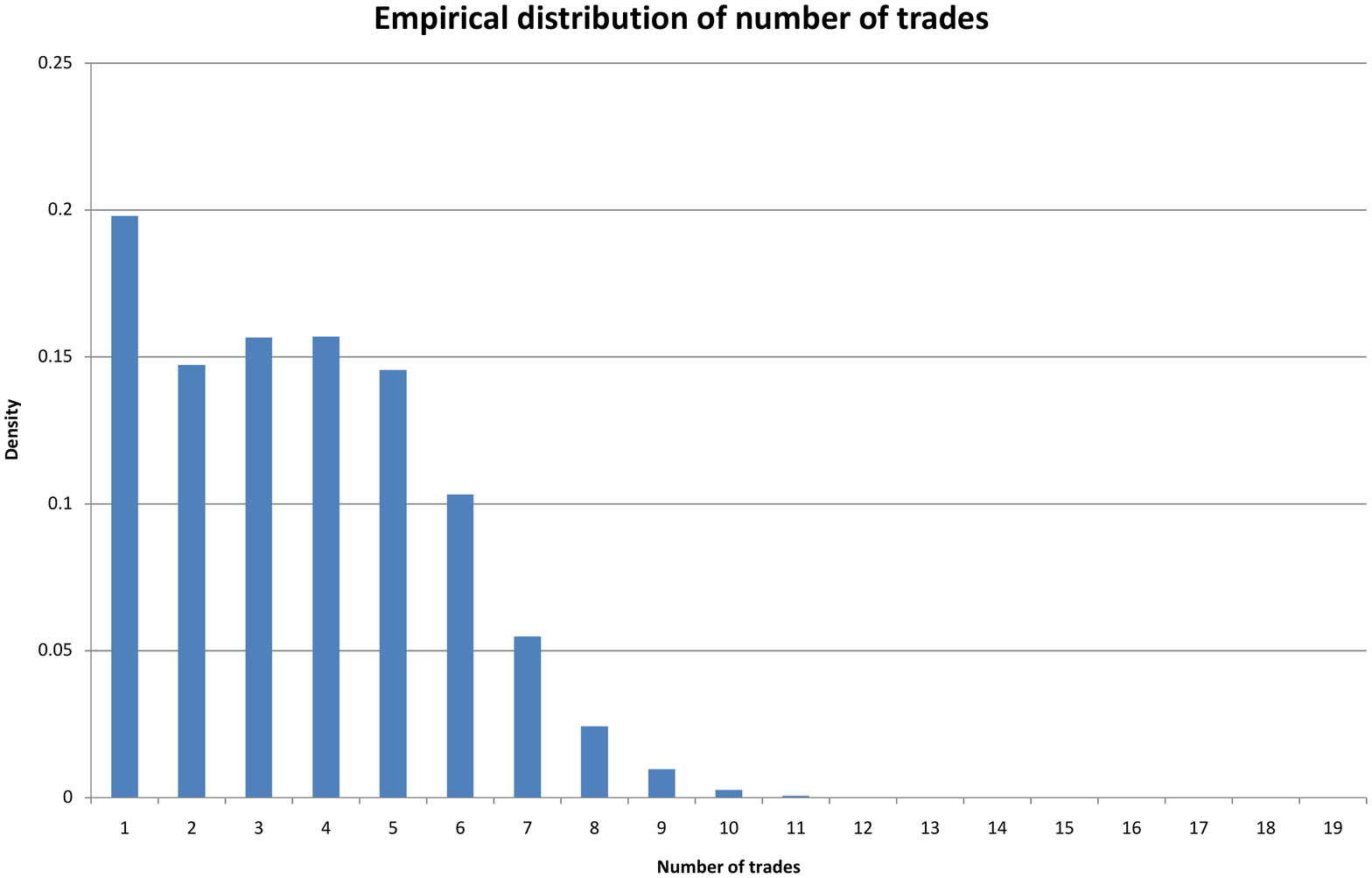}
\caption{Test 2: Empirical distribution of the number of trades }
\label{T2empdistribnumtrades}
\end{figure}

In this paragraph, we analyze the behaviour of the strategy as follows: first, we plotted in figure \ref{T2empdistribnumtrades} the empirical distribution of the number of trades for one trading session. Secondly, we plotted trades realizations for three days of the BNPP.PA stock for the week starting on 
04/19/2010. 

The three following graphs represent three days of market data for which we computed the mid-price (lines) with associated trades realizations for the optimal strategy (vertical bars). A positive quantity for the vertical bar means a buying operation, while a negative quantity means a selling operation.

\begin{figure}[h!]
\centering
\includegraphics[width=0.9\textwidth]{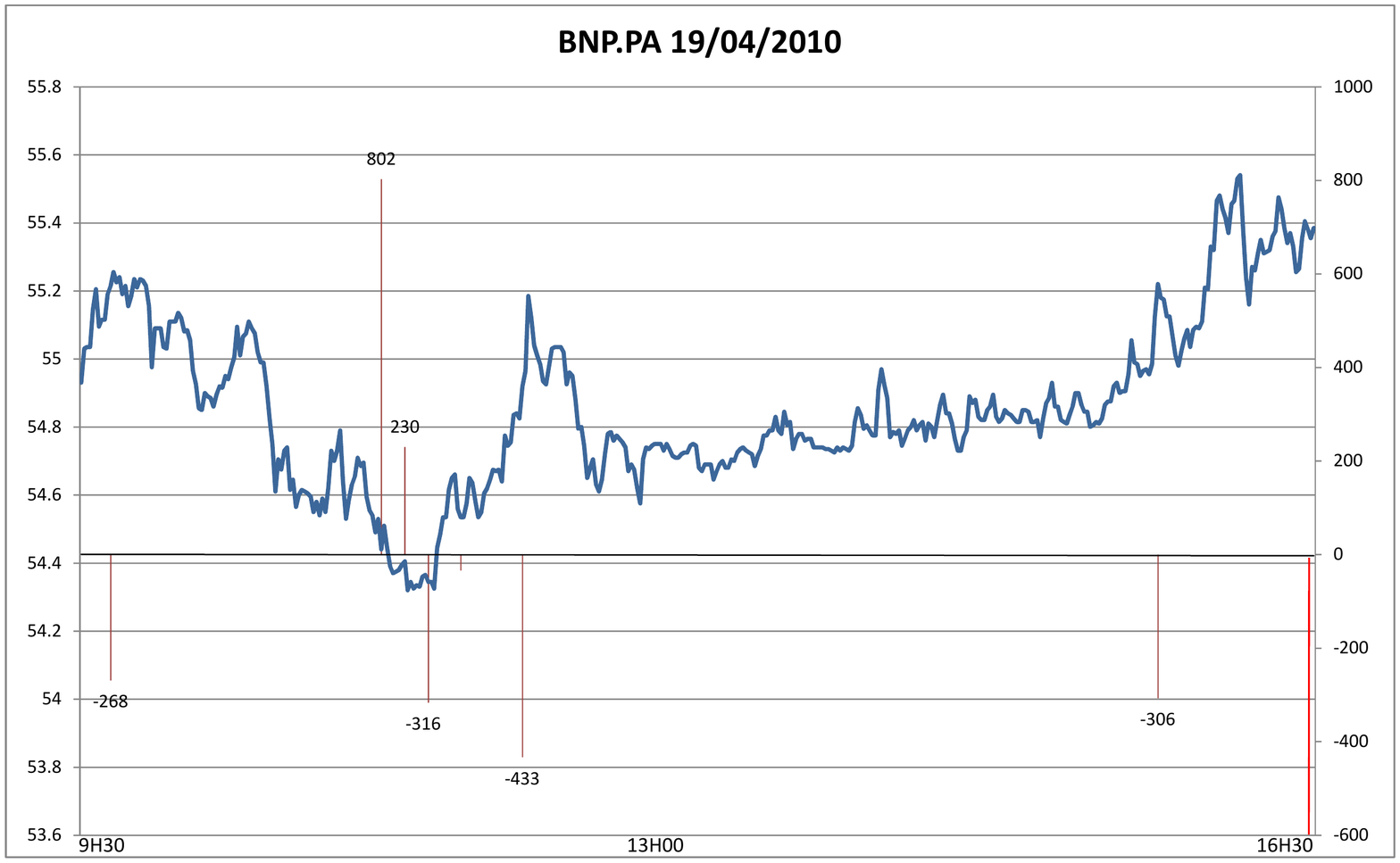} 
\caption{Test 2: Strategy realization on the BNP.PA stock the 04/19/2010. }
\label{T2BNP19}
\end{figure}
Figure \ref{T2BNP19} shows the trade realizations of the optimal strategy for the day 04/19/2010 on the BNPP.PA stock. The interesting feature in this first graph is that we see two buying decisions when the price goes down through the 54.5 $\geneuro{}$ barrier, and which corresponds roughly to a daily minimum. The following selling decision can be viewed as a failure. On the contrary, the two last selling decisions correspond quite precisely to local maxima. 
\begin{figure}[h!]
\centering
\includegraphics[width=0.9\textwidth]{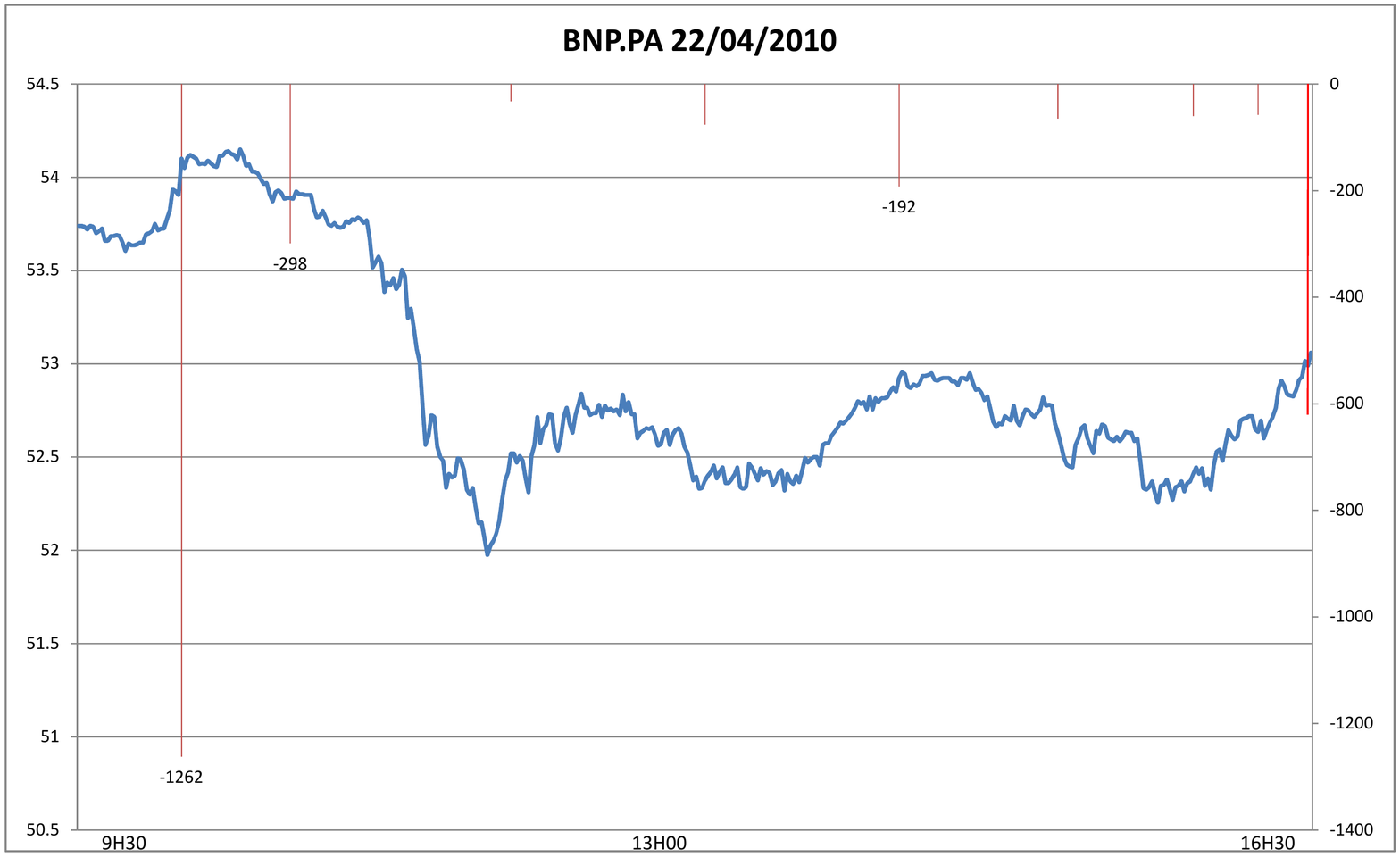}
\caption{Test 2: Strategy realization on the BNP.PA stock the 04/22/2010. }
\label{T2BNP22}
\end{figure}

Figure \ref{T2BNP22} shows the trade realizations of the optimal strategy for the day 04/22/2010 on the BNPP.PA stock. The interesting feature in this realization is that it looks like a U-shaped pattern of liquidation that appears in \cite{alfschsly09} for an mean-variance optimal strategy with a power-law order book resilience. This pattern is very robust, so we expect that it appears frequently in the optimal strategy. Moreover, we can notice that the volume of each trade in the day is roughly increasing (in absolute value) with the current price.

\begin{figure}[h!]
\centering
\includegraphics[width=0.9\textwidth]{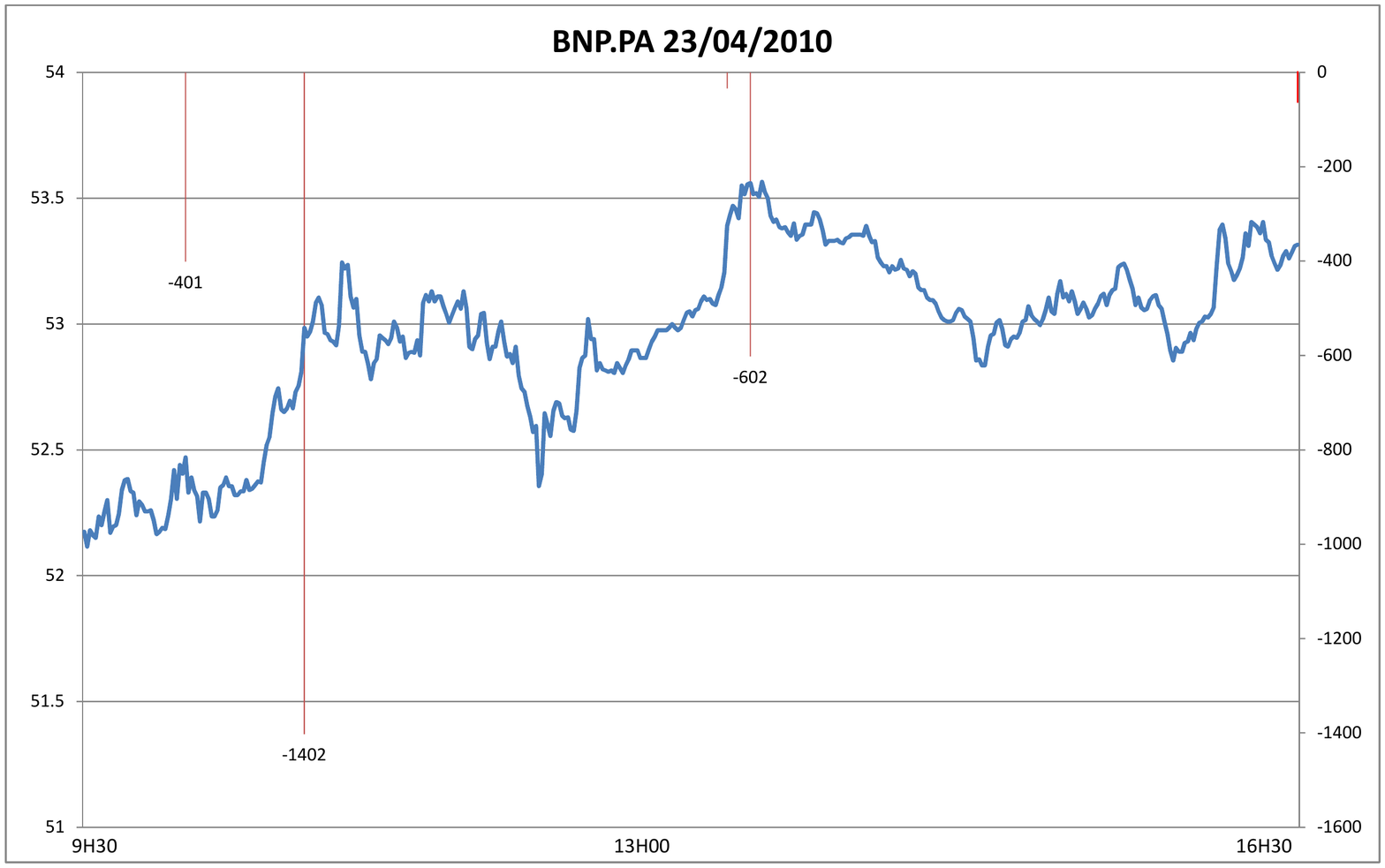}
\caption{Test 2: Strategy realization on the BNP.PA stock the 04/23/2010. }
\label{T2BNP23}
\end{figure}

Figure \ref{T2BNP23} shows the trade realizations of the optimal strategy for the day 04/23/2010 on the BNPP.PA stock. This realization is another illustration of the phenomenon of ``maxima detection" that appears when executing the optimal strategy. Note that in this last realization, the naive strategy was overperforming the optimal strategy, due to an unexpected price increase. Despite this, it is satisfactory to see that there are only three trades, which is less than on April 19 and 22, 2010, and that the detection of favorable price is accurate. 

\subsection{Test 3: Sensitivity to Bid/Ask spread}

In this last section, we are interested in the sensitivity of the results to the bid/ask spread, determined here by the two parameters $\kappa_a$ and 
$\kappa_b$. More precisely, we look at the dominant effect between the spread and the multiplicative price impact through the parameter $\lambda$. We proceeded to two tests here: one without bid/ask spread, i.e. $\kappa_a = \kappa_b =1$ and with $\lambda=5.10^{-4}$ as before, and one with a spread of $0.2\%$ and a price impact parameter $\lambda=0$.

\paragraph{Parameters} The table \ref{T3PARAM} shows the parameters of the two tests. We only changed the impact and spread parameters and let the others be  identical. 
\begin{small}
\begin{table}[h!]
\begin{center}
\begin{tabular}{|lll|lll|}
\hline Parameter & No spread test & No impact test & Parameter & No spread test & No impact test \\ 
\hline Maturity &	1 Day &	1 Day & $X_0$	&20000	&20000\\
$\lambda$	& 5.00E-04	& 0 & $Y_0$	&2500	&2500\\
$\beta$	&0.2&	0 & $P_0$	&51	& 51\\
$\gamma$ &	0.5	&0.5& $x_{min}$&	-20000	&-20000\\
$\kappa_a$ &	1	&1.001 & $x_{max}$ &	200000	&200000\\
$\kappa_b$	&1	&0.999 & $y_{min}$&	0	&0\\
$\epsilon$ &	0.001	&0.001 & $y_{max} $	&5000	&5000\\
$b$&	0.01	&0.01 & $p_{min}$	&49	&49\\
$\sigma$	&0.25	&0.25 & $p_{max}$	&53	&53\\
 & & & $n$	& 30	&	30\\
& & & $m$	& 40	&	40\\
& & & $N$	& 100	&	100\\
& &  & $Q$  & $10^5$& $10^5$\\
\hline
\end{tabular}
\end{center}
\caption{Test 3: Parameters}
\label{T3PARAM}
\end{table}
\end{small}

\begin{figure}
  \centering
  \includegraphics[width=0.9\textwidth]{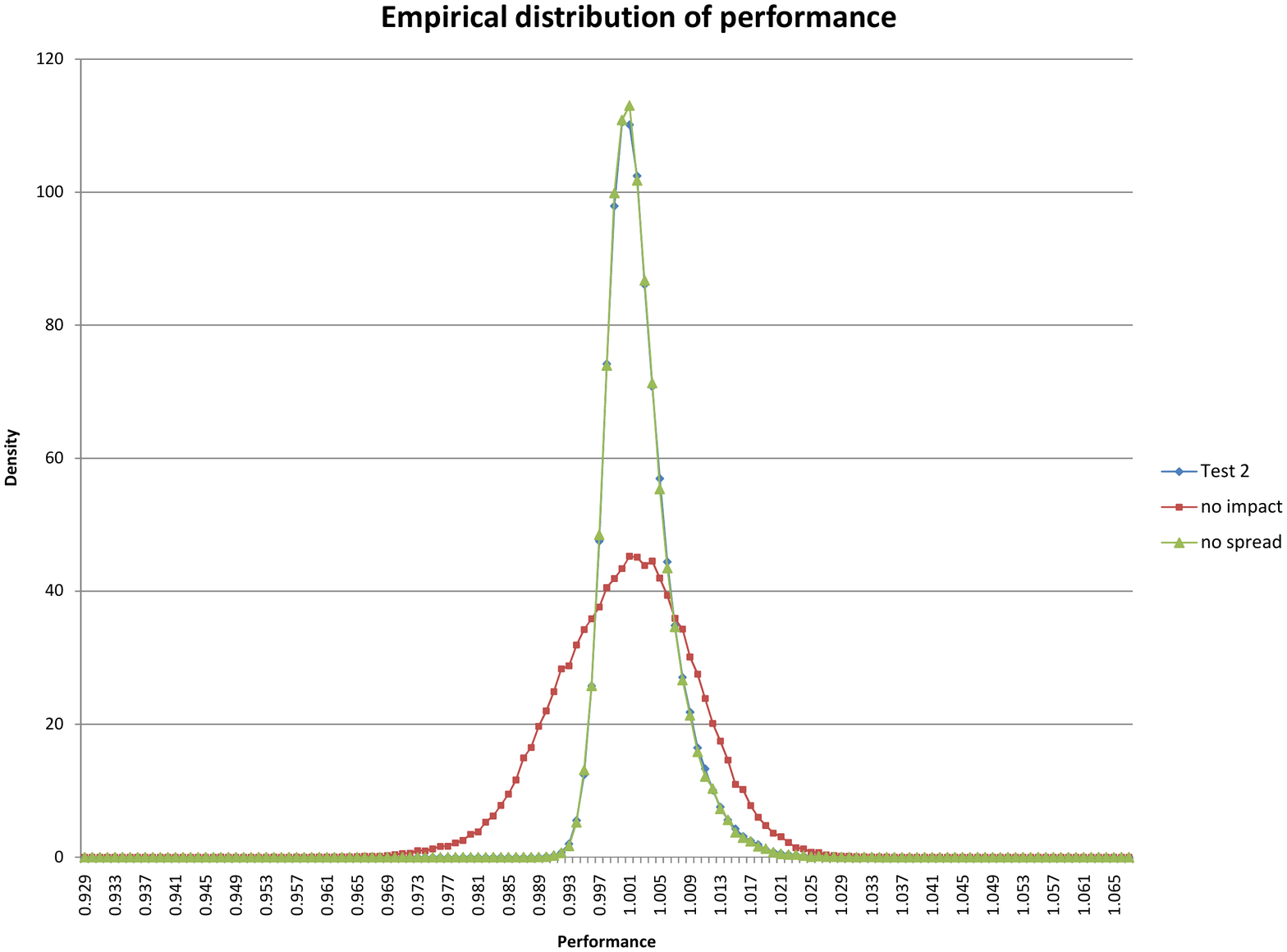} 
  \caption{Test 3: Empirical distributions of performance}
  \label{T3EMPDIS}
\end{figure}

\begin{figure}
  \centering
  \subfloat[No spread]{\label{fig:t3empdisnumtradwospread}\includegraphics[width=0.5\textwidth]{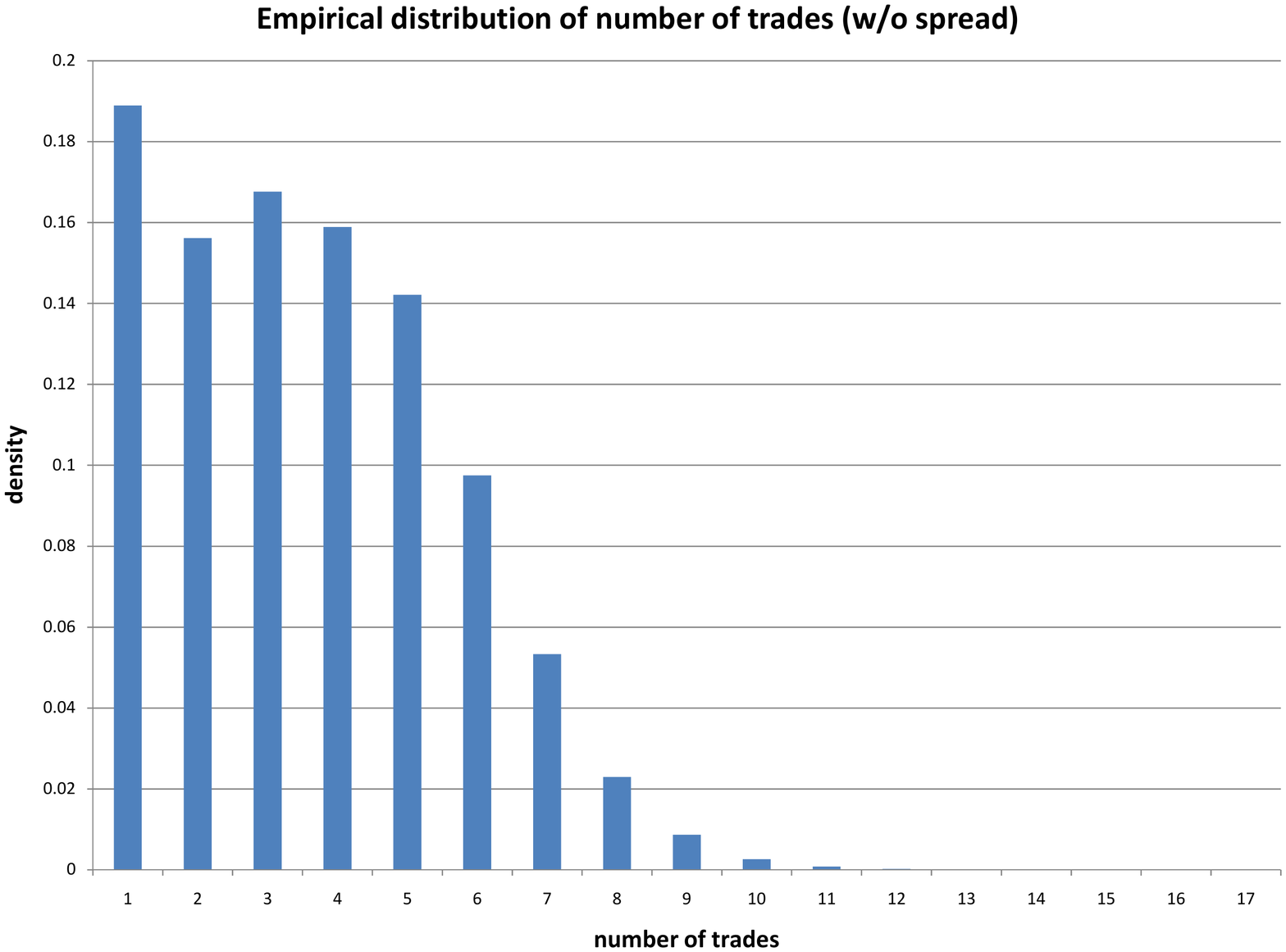}}                
  \subfloat[No impact]{\label{fig:t3empdisnumtradwoimpact}\includegraphics[width=0.5\textwidth]{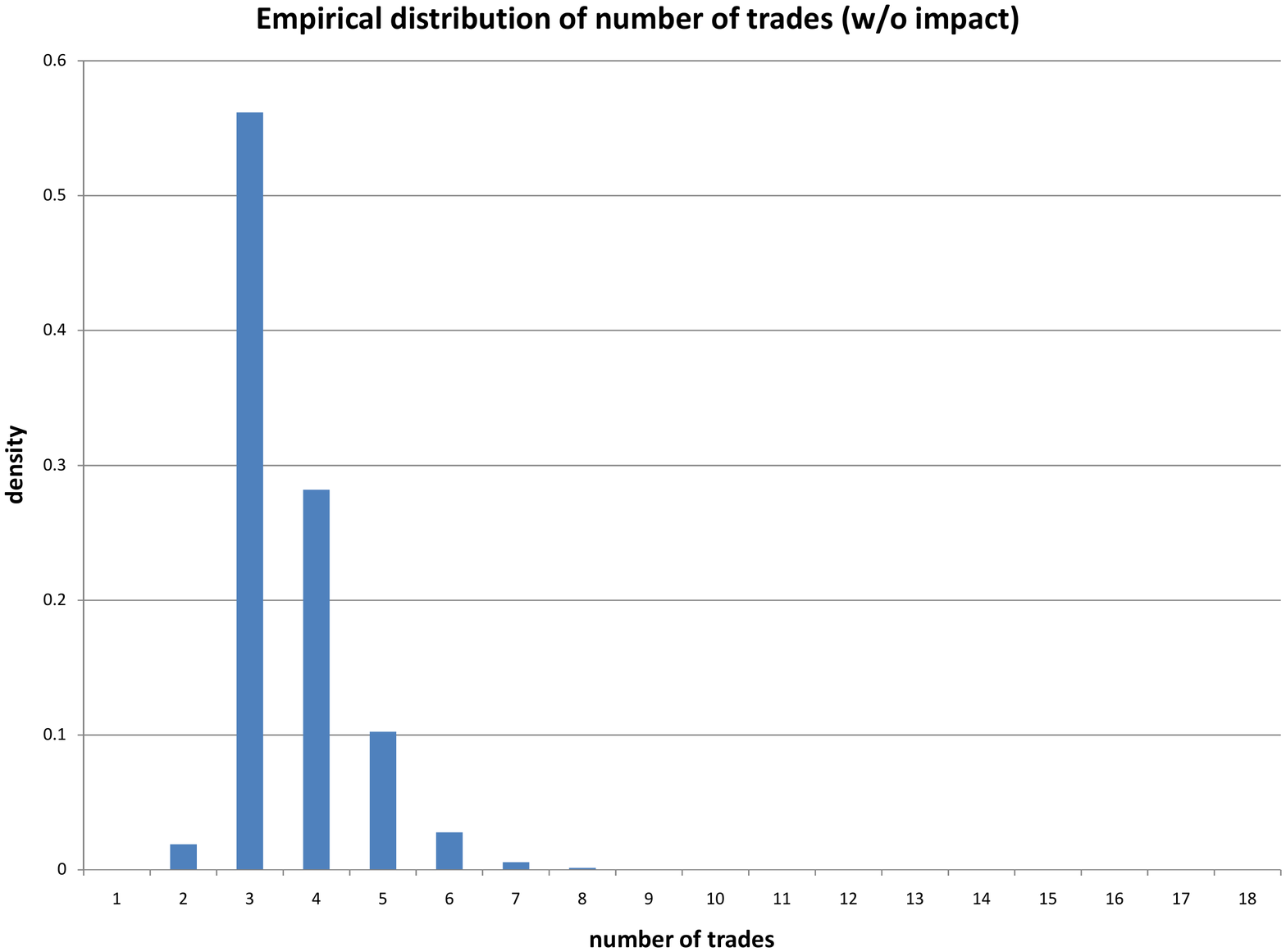}}
  \caption{Test 3: Empirical distributions of number of trades}
  \label{T3EMPDISNUMTRADES}
\end{figure}

\paragraph{Performance Analysis} In table \ref{T3Results} we computed several statistics on the results. In figure \ref{T3EMPDIS} we plotted the empirical distribution of performance in the two tests, with the test 2 distribution (Cf. figure \ref{T2empdistrib}) serving as a reference. In figure \ref{T3EMPDISNUMTRADES} we plotted the empirical distribution of the number of trades in the two tests, which is particularly helpful for interpreting 
the results.  Indeed, we see that a big spread reduces the number of trades of the optimal strategy. This can be interpreted as a phenomenon of clustering, which is consistent with the financial interpretation: if the spread is very large, the opportunity to buy at low price and sell at high price is significantly reduced, and more risky. Moreover, a large spread penalizes strategies that both buy or sell frequently. Then the optimal strategy tends to execute bigger quantity in a single trade, supported by the absence of  price impact. This has the side effect of enlarging the distribution of the optimal strategy in this case, since the number of trades tends to decrease. Indeed, we observe  that the smallest standard deviation is obtained with the uniform strategy, and the largest one is obtained with the naive strategy, then, qualitatively speaking, we expect the standard deviation of performance to  decrease with the number of trades.
On the other hand, setting the spread to zero does not change the shape of the empirical distribution, and comparing table \ref{T3Results} with tables \ref{T2PERF} and \ref{otherstats}, we see that there is almost no change between zero spread and a one-tick spread. This result may be interpreted  as follows: first, we  notice that there is an approximation in the state space, and particularly in the price grid, due to our discretization, which  is bigger than the scale of one tick (one tick is the price discretization unit in the market's limit order book, typically 0.01\geneuro{}). One finer study would be to set the prices grid precisely on the market's prices grid. However, as mentioned before, the special feature of our strategy is the path-dependency. Let us consider the typical scale of quantities involved in our optimization: we expect the optimal strategy to profit from price variation at the scale of 1\geneuro{} in our example; if the spread is about 0.1\geneuro{}, like in our last example, and if we usually do about 10 trades on the liquidation period, the effect of the spread ($10 \times 0.1$ \geneuro{}= 1\geneuro{}) is at the same scale as the price fluctuation. Then, the spread will have an important penalizing impact on the optimal trading strategy, and particularly on its schedule, i.e. the trading times (the quantities traded are more specifically constrained by market impact). On the contrary, if the spread remains small compared to the price fluctuations, the optimal trading schedule is not really modified, and the aspect of the performance distribution has a similar form. 

\begin{small}
\begin{table}[h!]
\begin{center}
\begin{tabular}{lllll}
\hline Quantity  & No spread test & No impact test & No spread vs. T2 & No impact vs. T2 \\ 
\hline 
Mean Utility  &	1.00113	& 1.00025 & $-3.00.10^{-5}$ & $-9.08.10^{-4}$ \\
Mean Performance  &	1.00227 &	1.00053 & $-5.98.10^{-5}$ & $-1.80.10^{-3}$ \\
Standard Deviation  &	0.00432 &	0.00906 & $-9.17.10^{-3}$ & $1.078$ \\
\hline
\end{tabular}
\end{center}
\caption{Test 3: Statistics. In the two last columns ''No spread vs. T2" (resp.''No impact vs. T2") are shown the relative values of ''No spread" test (resp. ''No impact" test) against the values of test 2 of the preceding section.}
\label{T3Results}
\end{table}
\end{small}

\newpage

\begin{small}

\end{small}

\end{document}